\documentclass[amsfonts, amssymb, amsmath, subscriptaddress, showpacs, reprint]{revtex4-1}
\usepackage{times}
\usepackage{mathrsfs}
\usepackage{amsbsy}
\usepackage[usenames,dvipsnames]{color}
\usepackage{bm}
\usepackage{graphicx}
\usepackage{color}

\begin{document}

\title{A nonlinear approach to transition in subcritical plasmas with sheared flow}
\author{Chris C. T. Pringle} 
\email{chris.pringle@coventry.ac.uk}
\affiliation{Applied Mathematics Research Centre, Coventry University, Coventry, CV1 5FB, United Kingdom}
\author{Ben F. McMillan}
\email{b.f.mcmillan@warwick.ac.uk}
\affiliation{Department of Physics, Centre for Fusion, Space and Astrophysics, University of Warwick, CV4 7AL Coventry, United Kingdom} 
\author{Bogdan Teaca}
\email{bogdan.teaca@coventry.ac.uk}
\affiliation{Applied Mathematics Research Centre, Coventry University, Coventry, CV1 5FB, United Kingdom }
\begin{abstract}
In many plasma systems, introducing a small background shear flow is enough to stabilize the system linearly. The nonlinear dynamics are much less sensitive to sheared flows than the average linear growthrates, and very small amplitude perturbations can lead to sustained turbulence. We explore the general problem of characterizing how and when the transition from near-laminar states to sustained turbulence occurs; a model of the interchange instability being used as a concrete example. These questions are fundamentally nonlinear, and the answers must go beyond the linear transient amplification of small perturbations. Two methods that account for nonlinear interactions are therefore explored here. The first method explored is edge tracking, which identifies the boundary between the basins of attraction of the laminar and turbulent states. Here, the edge is found to be structured around an exact, localized, traveling wave solution; a solution that is qualitatively similar to avalanche-like bursts seen in the turbulent regime. The second method is an application of nonlinear, non-modal stability theory which allows us to identify the smallest disturbances which can trigger turbulence (the minimal seed for the problem) and hence to quantify how stable the laminar regime is. The results obtained from these fully nonlinear methods provides confidence in the derivation of a semi-analytic approximation for the minimal seed. 
\end{abstract}
%
%
\maketitle
\onecolumngrid

\section{Introduction} 

In plasma physics, as in neutral fluids, configurations exist which are linearly stable but allow long-lived turbulence to develop once a large enough displacement away from the laminar state is given.
These are referred to as subcritical configurations \cite{casson, mcmillan09, Roach09, vanwyck, Johnson2005,friedman15} and can play an important role in the analysis and engineering of transport in plasmas. As such, methods that characterize the near-laminar and turbulent states, their interface and the threshold required for the transition between states are valuable.
The methods presented here to analyze the transition from a quiescent (laminar) state to a turbulent state are not specific to any particular scenario, but will be illustrated by examining a class of subcritical systems in which a pressure-gradient driven instability is suppressed by sheared flows. This scenario has instances in non-trivial situations such as in tokamak physics, where temperature-gradient driven drift instabilities are suppressed by poloidal zonal flows or system-scale toroidal flows. However, this scenario occurs more generally in non-stationary fluids such as accretion disks.

The dynamics of sheared flows in plasmas are somewhat surprising. On one hand, even a vanishingly small shear is enough to completely linearly stabilize the system. On the other hand, the properties of turbulence tend to vary smoothly as flow shear is imposed, as long as the system is initialized far from the laminar state. The connection between linear and nonlinear properties of these systems is therefore unclear. For example, although these systems are formally linearly stable, in the sense that linear perturbations decay at late time, perturbation energies can still grow by orders of magnitude in these regimes\cite{WaltzTrans, Johnson2005, Squire2014} before being damped eventually. Conversely, a highly localized small perturbation may still be sufficient to push the plasma into the regime where nonlinear effects are important and permit the system to evolve towards sustained turbulence. The threshold perturbation amplitude depends on both the linear dynamics of the system and the nonlinear processes that balance the linear decay that would otherwise take place. Practically, a subcritical system with a large nonlinear instability threshold might be able to remain in a low transport state, but one with a small threshold might in practice always exhibit turbulence. This kind of consideration is taken into account in engineering pipes for transporting fluid, and may be required for predicting the behavior of subcritical plasmas. 

To fully understand the dynamics of subcritical plasmas we need to move beyond linear analysis, i.e. linear transient amplification of small perturbations, and address these issue from the perspective of the full nonlinear problem.
Previously, some subcritical tokamak plasma papers have looked at identifying a boundary in parameter space above which turbulence 
is permanent and below which it is transient (or even non-existent) in nature.\cite{Highcock,VanWyk2017} These works performed 
parameter scans of simulations started at very large initial amplitude and run for a determined period 
of time. We will approach the problem in the opposite direction. For fixed parameters, we consider the initial 
evolution of a smoothly varied range of 
initial conditions to separate those which lead to turbulence from those which relaminarize, an idea touched on previously in plasma research\cite{VanWyk2017,friedman15}.
This enables us to identify the basins of attraction of the laminar and turbulent states and how these basins change as the 
parameters are changed.

While the laminar state itself is fully identified by the linearized system, identifying the extent of its basin of attraction is a nonlinear problem. Moreover, analyzing the type and properties of the solutions located on the boundary separating these states can offer insight into the transition to turbulence, providing an understanding of the processes leading to self-sustained turbulence. The states that occur on this boundary are often considerably simpler than the fully developed turbulent state, and in the model problem, a simple traveling wave is found on the edge, allowing a direct understanding of how linear and nonlinear terms balance. Intriguingly, these traveling waves echo features of the
avalanche-like bursts that permeate the turbulent regime. These methods were designed to determine the threshold amplitude and shape of the smallest disturbance that triggers turbulence (i.e. the minimal seed), but the insight they offer into the dynamics of the transition to turbulence may well be of equal importance. 

In this paper we will present two fundamentally nonlinear methods of analyzing the transition to turbulence. We will do so in the context of a simple fluid interchange model (section \ref{sec:HW}), however they are general and applicable to any system with two (or more) linearly stable states. The first of these methods explores the boundary separating the basins of attraction of the two states (section \ref{sec:edge}). This boundary behaves in a simple but nonlinear manner and is dominated by a traveling wave solution (section \ref{sec:tw}). The second method we present looks at which disturbances are most dangerous in terms of triggering transition (section \ref{sec:nltg}). As a means to capture the generic parameter dependencies and provide insight into the the nonlinear growth process, a semi-analytical model of the minimal perturbation amplitude is derived (section \ref{sec:approx}). Both approaches illustrate that the critical dynamics are localized in space. This is in sharp contrast with the linear mode of maximal transient growth which is a space-filling plane wave. In section \ref{sec:conc} we draw these results together to describe an entire transition scenario.

\section{The Plasma Interchange model}\label{sec:HW}

The plasma model considered here, which we denote PI (for plasma interchange model), is a fluid equation with as effective gravity term giving rise to an instability: it was introduced in \citet{Beyer_PRE_61_813} and previously considered in \citet{mcmillan09} in order to explain the avalanches that arise in simulations of tokamak turbulence with a background shear flow. The equations are also similar in form to those encountered in local models of convection in gravitationally confined rotating plasmas (often subject to instabilities like the magneto-rotational instability)\cite{Johnson2005,Squire2014}.
It is closely related to the Hasegawa-Wakatani model, differing only in the linear terms that couple the density field to the temperature evolution. Physically, the instability arising in Hasagawa-Wakatani is of resistive-drift type, whereas the model here is an interchange instability. There are a wide class of `local' plasma models such as gyrofluid or gyrokinetic models, with instabilities driven by field line curvature or gravity, and nonlinearities due to plasma advection, and much of the nonlinear behavior, such as the formation of zonal flows, is generic. We therefore claim that this model will represent some of the generic features of plasma turbulence in the presence of a uniform drive from curvature or an external force such as gravity, and a background shear flow. The model is formally derived by taking a local limit of the MHD equations and parameterizing the dynamics along the field line, so that the resulting degrees of freedom describe plasma motion in terms of quantities in the 2D plane perpendicular to the field. 

The model evolves the fluid vorticity $\nabla^2 \phi$ (as in generic fluid equations such as Navier-Stokes), which is self-advected, and the fluid density $n$, which is advected by the velocity field. There is a gravitational force (representing the effective force due to magnetic field curvature) which induces a force on the fluid proportional to the density. As noted by Ref. \cite{Benkadda}, the 2D model equations are equivalent to Rayleigh-Bernard convection equations, but unlike in the Rayleigh-Bernard problem, where the fluid is confined between two plates, the domain here is 2D and infinite. Because we have added a uniformly sheared background flow, this is similar to what is sometimes referred to as the Rayleigh-Bernard-Couette (RBC) model. In the RBC model the most unstable modes are streamwise-uniform, that is, with wavevectors perpendicular both to the background flow and the gradient, but this direction is perpendicular to the 2D plane chosen: the high effective stiffness of magnetic field lines (which is not explicitly modeled) ensures that the motion is spanwise-uniform and justifies the use of a 2D model. This model is therefore a generic representation of interplay between a gradient driven instability (Rayleigh-Taylor) and shear flows, but of a slightly different form to the classic fluid models. 
The model may be written
\begin{equation}
 \frac{d}{dt} \nabla^2 \phi = \frac{g}{n} \frac{ \partial n}{\partial y} + \nabla^2 \phi
\end{equation}
\begin{equation}
  \frac{d}{dt} n   = \nabla^2 n
\end{equation}
where the fields $n$ and $\phi$ are functions of spatial coordinates $x$ and $y$ as well as time $t$, and $d/dt = \partial/\partial t - \nabla \phi \times \hat{z} . \nabla$ is the convective derivative.

To make this model applicable to microinstabilities in plasma, where only a narrow range of wavenumbers $k_y$ are actually unstable, we follow references \onlinecite{Benkadda,mcmillan09} and consider only the modes $k_y=0$ (which represents a zonal flow or density perturbation) and $k_y=k_{y0}$ (representing a drift mode). This reduces the spatial dimensionality to 1D and is appropriate where the dynamics are dominated by a narrow range of unstable modes and coupling to the zonal flow. This is motivated by the dominant role of zonal perturbations in tokamak turbulence dynamics and by post-hoc qualitative similarity of the features of this model to gyrokinetic simulations\cite{mcmillan09}. The model captures the linear drift wave instability process, zonal flow generation via Reynolds stresses and shear flow stabilization of turbulence by zonal flows. Normally, system-scale perturbations (at low $k_y$) would dominate in the Rayleigh-Bernard model, but these are absent by construction as only a single non-zero $k_y$ is retained (physically, the reason that longer perpendicular wavelength mode are stable for electrostatic drift instabilities is because Landau damping dominates over charge separation). The short wavelength compared to the system scale justifies the use of a local treatment with periodic boundary conditions. 

The model comprises of four 1d equations corresponding to the zonal averaged density field ($\bar{n}$), the 
gradient of the zonal electric potential ($E$) and the complex wave potential ($\tilde{\phi})$ and wave density 
($\tilde{n}$). These quantities evolve according to the equations
\begin{align}
\partial_t\bar{n}&=i\partial_x(\tilde{n}^*\tilde{\phi}-\tilde{n}\tilde{\phi}^*) \label{eqn:evol1} \\
\partial_tE&=i\partial_x(\tilde{\phi}\partial_x\tilde{\phi}^*-\tilde{\phi}^*\partial_x\tilde{\phi})+\partial_{xx}E \\
\partial_t\tilde{\phi}&=-i\tilde{\phi}(Sx+E) +i\tilde{n}+\partial_{xx}\tilde{\phi}  \\
\partial_t\tilde{n}&=-i\tilde{n}(Sx+E) +i\tilde{\phi}(\partial_x \bar{n} -1) +\partial_{xx}\tilde{n}, \label{eqn:evol2}
\end{align}
where $S$ is the background shear. These are solved in the domain $0\leq x \leq L$ subject to the boundary conditions
$\bar{n}(L)=\bar{n}(0)$, $E(L)=E(0)$, $\tilde{\phi}(L)=\tilde{\phi}(0)\exp\{-itLS\}$ and 
$\tilde{n}(L)=\tilde{n}(0)\exp\{-itLS\}$. Diffusion on the zonal density $\bar{n}$ was been removed on the principle that
a collisionless plasma was being simulated; it was nevertheless retained for $\bar{\phi}$ because the long wavelength
zonal flows otherwise build up excessively (compared to the gyrokinetic simulations), possibly because the tertiary
instability is missing. In a collisionless setting, the diffusion term on the drift modes may be thought of as arising
due to a $k_x$ dependent growth rate.

The behavior of this system is explored in reference \onlinecite{mcmillan09}: note that scaling transformations have been used to reduce the number of free parameters. In the absence of a background flow ($S=0$), the linear dynamics may be captured straightforwardly in terms of a collection of non-interacting plane waves growing due to the interchange instability, but damped due to the diffusion term, so the growth rate is maximum when the wavenumber $k_x=0$. However, for $S>0$ initially plane wave modes become tilted due to the sheared background flow (in the 2D plane), and the wavenumber $k_x$ increases linearly with time. Given any initial plane wave perturbation, the late-time wavenumber $k_x$ will eventually increase until the mode is damped. The system is thus linearly unstable for $S=0$, but stable for all positive values of $|S|$.
This situation has a familiar spatial analogue, the convective instability, where a mode propagates through a finite unstable region, and is amplified for a certain time, but later propagates into a stable region, where it is damped. 

\begin{figure}
\begin{center} 
\resizebox{0.9\textwidth}{!}{\includegraphics[angle=0]{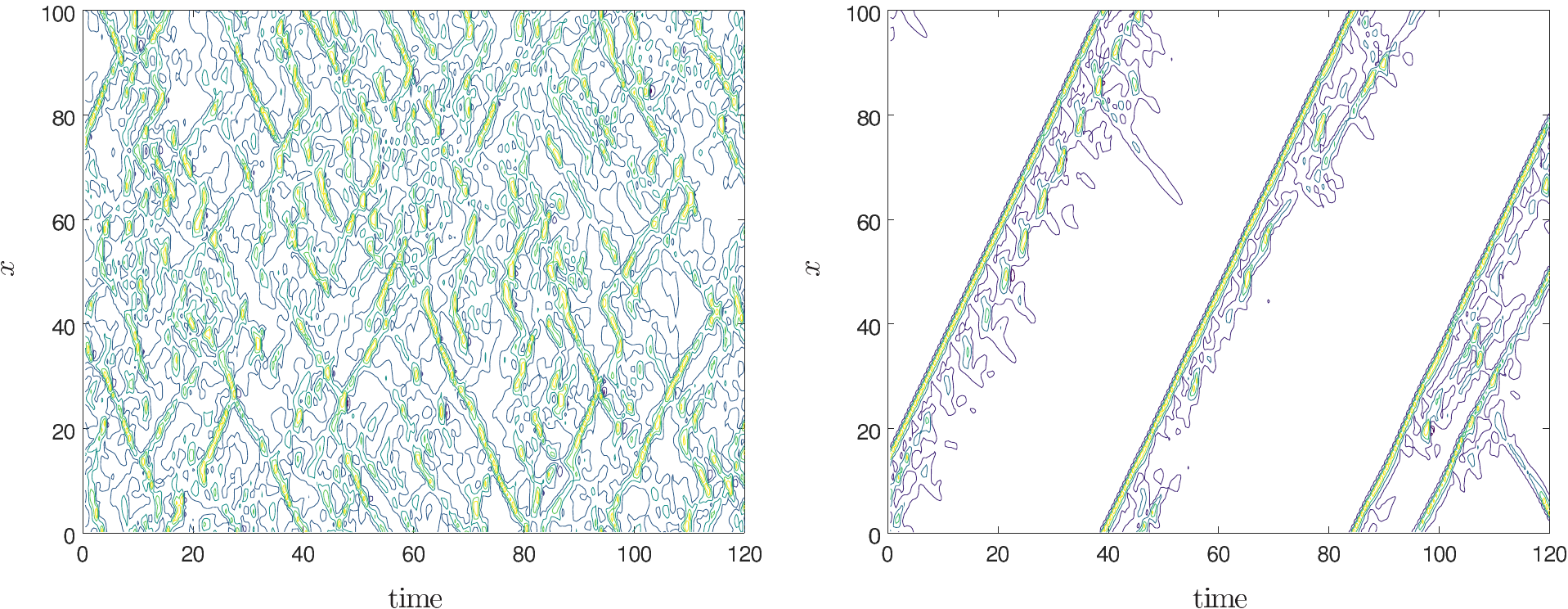}}
\end{center}
\caption{\textbf{Left}: Contours of flux, $Q=\tilde{n}^*\tilde{\phi}-\tilde{n}\tilde{\phi}^*$, on a plot of space 
against time for the PI model for $S=0.15$. The turbulence is chaotic and fills the domain, but 
not without structure as individual sub-patches advect across the domain.
\textbf{Right}: Again contours of flux but this time with $S=1.1$. Now the turbulence is localized and travels 
across the domain at a constant rate leaving small fluctuations in its wake.}
\label{fig:hwturb}
\end{figure}

Despite the linear stability, chaotic turbulent flow is observed at finite values of $S$. For values of the shear below $S\approx 0.5$ the turbulence fills the domain (figure~\ref{fig:hwturb}, left), however above this value it becomes increasingly localized. This continues to the extent that for $S=1.1$, burst like solutions are identified traveling across the domain at near constant rate with only slow variations of small amplitude (figure \ref{fig:hwturb}, right). If the shear is increased past this point, turbulence becomes increasingly transient and ceases to be sustainable. 

\section{The edge of chaos}\label{sec:edge}

An important consequence of the linear stability of this system is that finite amplitude disturbances are required to trigger turbulence. In general, for any given disturbance there exists a critical amplitude below which 
the disturbance dissolves away and above which it leads to chaos. If the critical amplitude itself is chosen then 
the flow continues on in an inbetween state, never relaminarising and never becoming truly turbulent. This 
manifold is referred to as the edge of chaos (or laminar-turbulent boundary) and has been extensively studied in a range 
of subcritical shear flows.\cite{itanoToh01,skufca06} The edge of chaos itself has undergone more limited 
investigation in plasma models.\cite{friedman15}

The edge of chaos is the boundary between the basins of attraction for the laminar and turbulent states. In a 
phase space visualization of the system, the edge forms a hypersurface. Any initial 
condition `outside' or `above' the edge will lead to turbulence, while those below the edge will laminarise. 

We can locate the edge for the PI system by choosing a form of perturbation and iteratively refining 
its amplitude through bisection. If turbulence is observed the amplitude is reduced, while if the disturbance decays away the amplitude is increased. As the refinement increases we are able to track to initial conditions, one on either side of the edge. Initially their evolutions mirror each other, but eventually they begin to drift apart as one relaminarises and the other becomes turbulent. The more accurately the critical amplitude is determined the longer the separation of the two trajectories can be delayed. If the precise amplitude of the edge could be found the two trajectories would never separate. Instead, the limitations of numerical accuracy mean that divergence is inevitable. In order to be able to track the edge past this, the approach is restarted with the two states just after they begin to depart the edge by interpolating between them.

\begin{figure}
\begin{center} 
\resizebox{0.6\textwidth}{!}{\includegraphics[angle=0]{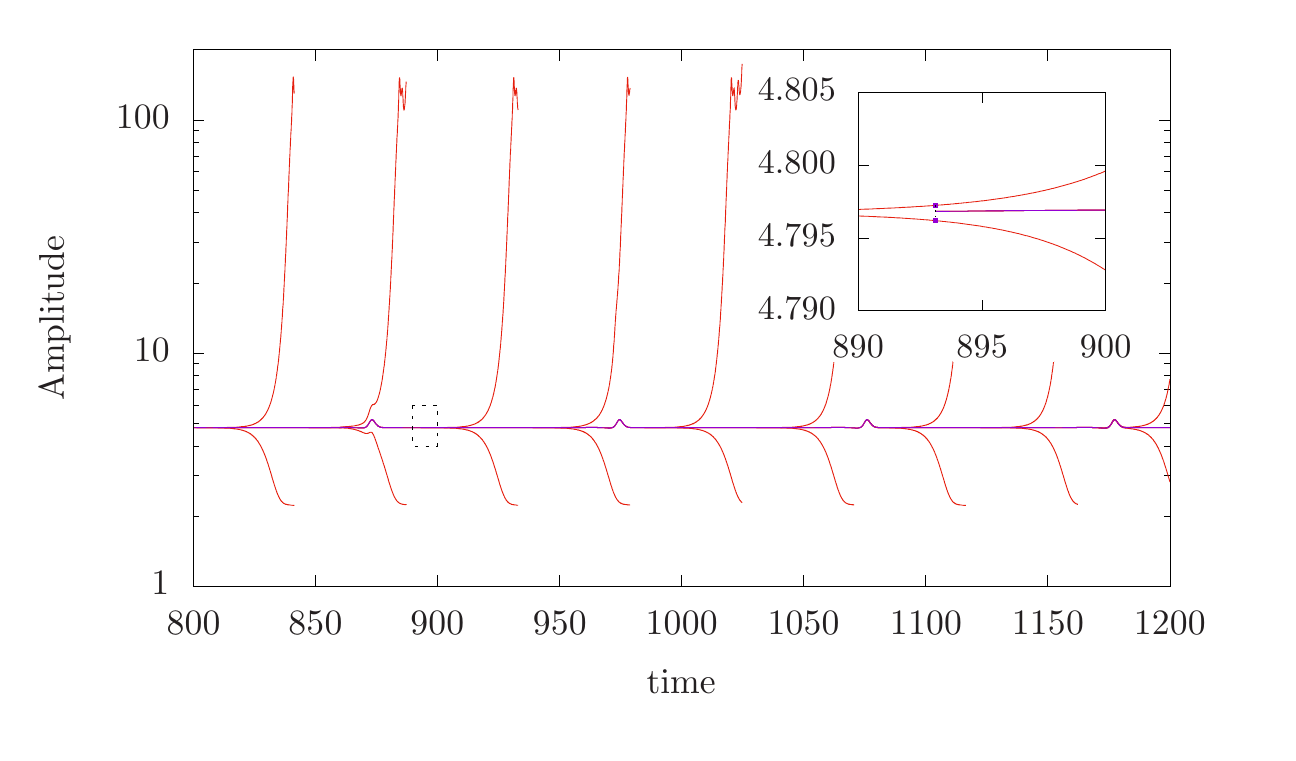}}
\end{center}
\caption{\textbf{Main}: Time evolution of energy for the edge of chaos. The edge was found by bisection 
and the calculation restarted when numerical precision failed.   \textbf{Inset}: A magnified version of 
one of the restart points. The states on either side of the edge can be clearly seen as can the interpolated 
state within the edge.}
\label{fig:edge}
\end{figure}

In figure \ref{fig:edge}, the results of the edge tracking calculation can be seen. After an initial transient, 
the simulation settles down. Due to a lack of diffusion, a background distribution of $\bar{n}$ develops. Superimposed
on top of this the dynamics become simple, an isolated traveling wave advecting at a constant rate. 

\section{The edge state}\label{sec:tw}

All initial conditions within the PI edge, for any given value of the background shear, evolve into 
the same state -- henceforth referred to as the edge state. This simple traveling wave is an exact solution 
of the underlying equations (exact in the sense that they satisfy the equations, propagating at a constant rate 
without variation or fluctuation), and fully nonlinear in nature. That all initial conditions within the edge evolve into this edge state is reflective of the fact that this state is linearly stable \emph{within the edge}. In an absolute 
sense it is unstable, however its only unstable direction points out of the edge so it does not affect the edge 
dynamics. The scenario of an attracting edge state has also been observed in a number of other systems, albeit 
usually only after restricting symmetries have been imposed.\cite{schneider08,dwk08}

The traveling wave can be calculated directly. It takes the form
\[
(\bar{n}(x,t),E(x,t),\tilde{\phi}(x,t),\tilde{n}(x,t))
=(\bar{n}(x-ct),E(x-ct),\tilde{\phi}(x-ct)e^{-icSt^2/2},\tilde{n}(x-ct)e^{-icSt^2/2})
\]
and so is a solution to the equations
\begin{align*}
\partial_t \bar{n}&=-c\partial_x \bar{n}\\
\partial_t E&=-c\partial_x E\\
\partial_t \tilde{\phi}&=-c\partial_x \tilde{\phi}-icSt\tilde{\phi} \\
\partial_t \tilde{n}& =-c\partial_x \tilde{n}-icSt\tilde{n}. 
\end{align*}
Equations (\ref{eqn:evol1})-(\ref{eqn:evol2}) are substituted into the lefthand side to form a coupled eigenvalue problem.
The equations are supplemented by applying a phase condition. 
We solve this by making an initial guess for the travelling wave. 
Solutions are then sought iteratively using a Newton-Raphson
approach. As this is a relatively small dynamical system (we allowed $3000$ degrees of freedom), the Jacobian can 
be explicitly calculated. As always, successful convergence of such an approach depends on making a suitable 
initial guess.  For this problem we are able to do so simply by taking the end state of the edge tracking. 

\begin{figure}
\begin{center} 
\resizebox{0.6\textwidth}{!}{\includegraphics[angle=0]{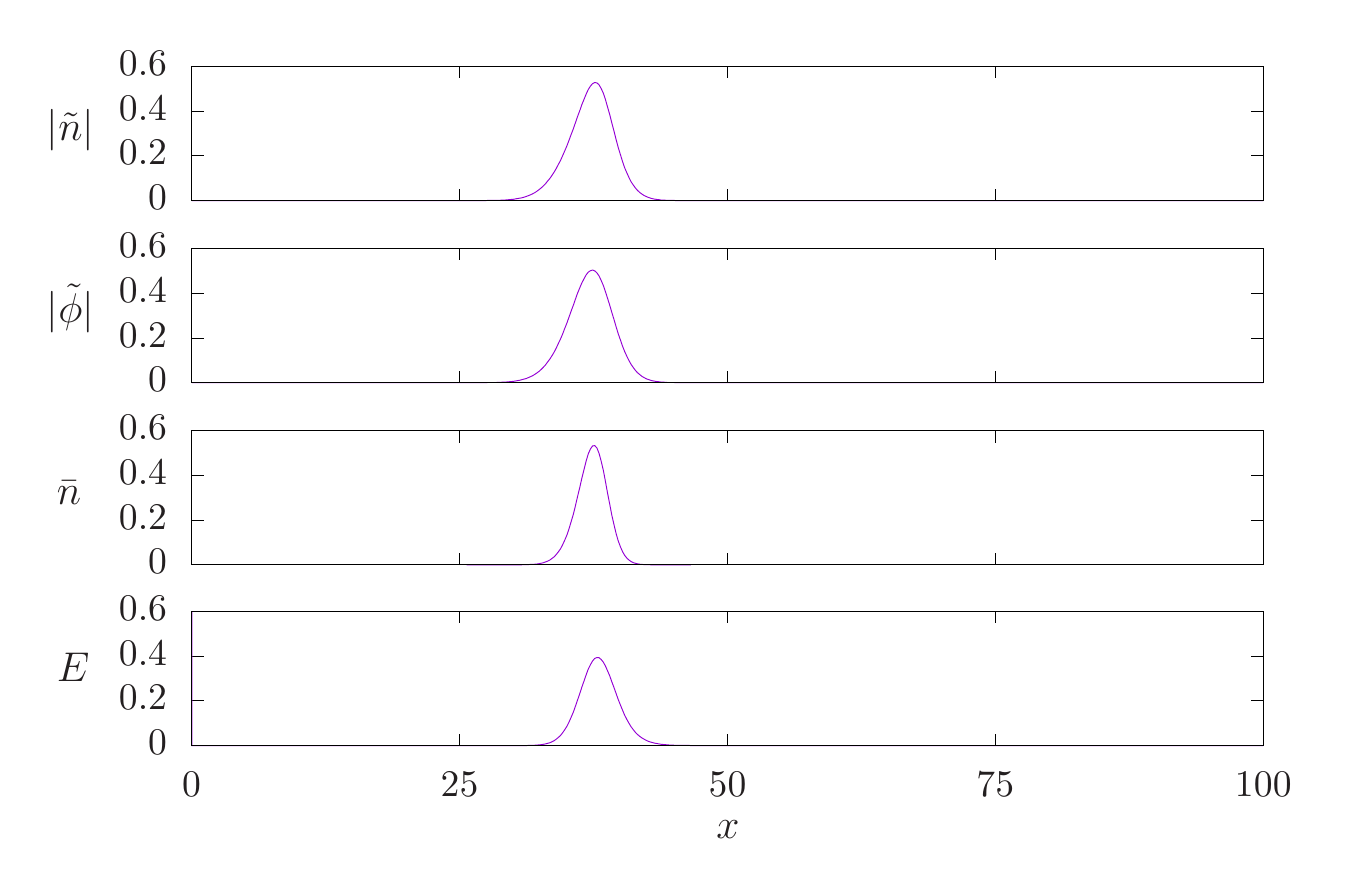}}
\end{center}
\caption{The exact traveling wave solution for $S=0.5$. The solution is clearly localized to one section 
of the domain. All four components contribute to the traveling wave and combine to sustain it against 
diffusive decay.}
\label{fig:tw}
\end{figure}

\begin{figure}
\begin{center} 
\resizebox{0.6\textwidth}{!}{\includegraphics[angle=0]{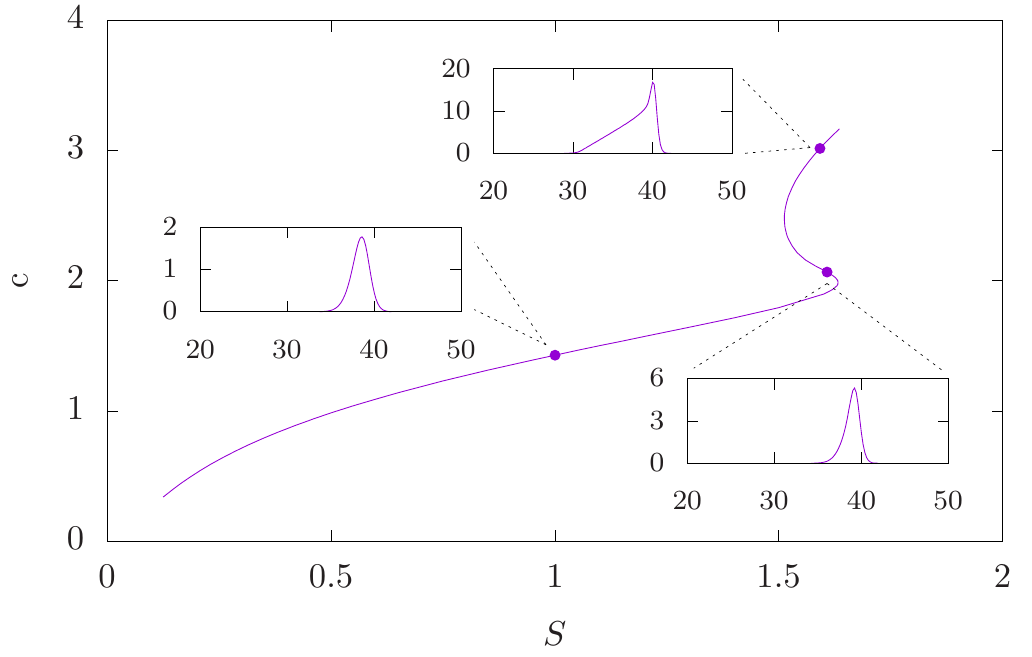}}
\end{center}
\caption{\textbf{Main:} Tracking the traveling wave solutions as the background shear is varied. The maximum shear rate it exists at is $S=1.6325$, beyond which it doubles back. \textbf{Inset:} Plots of $\bar{n}$ for the traveling wave solution at three different points. Only part of the domain is shown. As the traveling wave is traced upwards it increases in amplitude and the spatial interface becomes more abrupt.}
\label{fig:e_c}
\end{figure}

In figure \ref{fig:tw} we plot the traveling wave for $S=0.50$. It is fully localized (in the sense that the 
solution does not change when the length of the periodic domain is increased). Having found this one solution we can then explore how it varies as the background shear changes. Beginning from our initial solution we set $S$ to be slightly changed and then use our previous solution as the initial guess in our Newton-Raphson routine at the new shear rate. When we do so we are then able to track the solution through phase space, as seen in figure \ref{fig:e_c} (main). Increasing $S$ leads us to find there is a maximum shear rate ($S=1.6325$) for which this solution exists. Beyond this point the solution turns back. Having done so, rather than then continue back to zero shear, it then snakes upwards for as far as we continued it. During this progress, the form of the traveling wave changes drastically and its amplitude rapidly increases. It remains localized, but the spatial interface between the wave and the surrounding empty space becomes abrupt as shown in figure \ref{fig:e_c} (right). 

The traveling wave solution appears to be a natural counterpart to the propagating bursts seen in fully developed turbulent systems\cite{candy_d3d,mcmillan09,gene_bursts,vanwyck}: the amplitude, velocity and spatial profile of the propagating bursts in the turbulent manifold of the PI model\cite{mcmillan09} are quite similar to the edge state near the critical shear $S = 1.6325$. The states found in the turbulent phase, however, do not appear to be exact traveling waves, and may be periodic or relative periodic orbits: we made an attempt to find an exact traveling wave in the turbulent manifold but were not able to locate any.

\section{The minimal seed}\label{sec:nltg}

The traveling wave calculated in the previous section dominates the behavior within the edge. In this sense 
it represents the typical amplitude of disturbance required to trigger a turbulent episode. The actual amplitude required will depend on the precise form of perturbation applied. Some forms are more efficient than others and may trigger turbulence at much lower amplitudes. With this in mind it seems pertinent to ask, what is the smallest disturbance capable of triggering turbulence? It may be of practical importance to know what the minimal disturbance required in order to construct a `maximum safe level' of experimental fluctuation below which the system is guaranteed to remain laminar. Identifying this disturbance is equivalent to finding the point within the edge with the lowest amplitude. This state is called the minimal seed.\cite{pwk12}

To answer this, we begin by considering how disturbances inside the edge behave. We have already seen that for 
the PI model there is a simple attractor within the edge. The nature of this state means that for any 
initial condition chosen from within the edge, given enough time it will evolve into this simple traveling wave. 
The immediate implication of this is that no matter the initial amplitude of a disturbance within the edge, its 
eventual amplitude will be given by that of the traveling wave solution. If we denote a state, evolving in time, 
as $\mathbf{x}(t)$, then for all disturbances within the edge
\begin{equation}
\lim_{t\rightarrow\infty} \langle\mathbf{x}^2(t)\rangle = \langle \mathbf{x}_{TW}^2\rangle.
\end{equation}
where 
\begin{equation}
\langle \cdots \rangle =\int_0^L \cdots \,dx
\end{equation}
and $\mathbf{x}_{TW}$ is the traveling wave solution.

As a consequence of this, the minimal seed -- the disturbance within the edge with the smallest amplitude -- will also be the initial condition within the edge which maximizes the quantity
\begin{equation}
\lim_{t\rightarrow\infty} \frac{\langle\mathbf{x}^2(t)\rangle}{ \langle\mathbf{x}^2(0)\rangle} =
 \frac{\langle\mathbf{x}_{TW}^2\rangle}{ \langle\mathbf{x}^2(0)\rangle}.
\end{equation}
To identify this state we begin by considering the related question: of all 
possible disturbances of a given initial amplitude, which exhibits the most growth, $G$, over asymptotically large times
\begin{equation}
G(\mathbf{x}(0))= \lim_{t\rightarrow\infty} \frac{\langle\mathbf{x}^2(t)\rangle}
{\langle\mathbf{x}^2(0)\rangle}.
\end{equation}
If the initial amplitude is fixed to be that of the minimal seed, $A_0=A_{MS}$, then clearly $G$ will be maximized by the minimal seed itself. This can be seen as for the minimal seed $G=\langle\mathbf{x}_{TW}^2\rangle/
\langle\mathbf{x}_{MS}^2\rangle$ (by definition the minimal seed is within the edge and so must evolve 
into the traveling wave), while all other disturbances are below the edge and so relaminarise implying $G=0$ for them. Even if only large rather than infinite times are considered, the minimum seed should still maximize this quantity.

The approach is hampered by the fact that we do not \emph{a priori} know $A_{MS}$. Instead, if we maximize $G$ for a 
succession of increasing initial amplitudes we should be identify the point at which the edge is crossed by a sudden
increase in the amount of growth produced. How sudden this increase is will depend on how large we take $T$ to 
be.\cite{kpw_review} 

The challenge is now to identify those perturbations which generate the most growth. Small disturbances which can exhibit large growth have historically been the subject of the study of transient growth. For the linear problem, identifying the disturbances which grow the most was originally done by considering the pseudo-spectra of the operator. More recently the problem is frequently considered by formulating a functional to be maximized (see reference \onlinecite{schmid07} and references therein). The attraction of 
this approach is its flexibility and the fact that it can typically be solved using minor modifications to a standard 
time-stepping code. Importantly, it is straightforward to include the nonlinear terms in the calculation 
\cite{PK10, cherubini10, mono11}. 

Let us define
\begin{align}
\mathscr{L}=&\langle \bar{n}(x,T)^2+E(x,T)^2+\tilde{\phi}(x,T)^2+\tilde{n}(x,T)^2 \rangle \nonumber \\ 
&-\lambda\big(\langle \bar{n}(x,0)^2+E(x,0)^2+\tilde{\phi}(x,0)^2+\tilde{n}(x,0)^2 \rangle -A_0\big)\nonumber \\
&-\int_0^T \langle \bar{N}.\big(\partial_t\bar{n}-i\partial_x(\tilde{n}^*\tilde{\phi}-\tilde{n}\tilde{\phi}^*)\big) \rangle \,dt\nonumber \\
&-\int_0^T \langle F.\big(\partial_tE-i\partial_x(\tilde{\phi}\partial_x\tilde{\phi}^*-\tilde{\phi}^*\partial_x\tilde{\phi})-\partial_{xx}E\big)\rangle\,dt \nonumber \\
&-\int_0^T\langle\tilde{\Phi}^*.\big(\partial_t\tilde{\phi}+i\tilde{\phi}(Sx+E) -i\tilde{n}-\partial_{xx}\tilde{\phi}\big) \rangle \,dt\nonumber \\
&-\int_0^T\langle \tilde{N}^*.\big(\partial_t\tilde{n}+i\tilde{n}(Sx+E) -i\tilde{\phi}(\partial_x \bar{n} -1) -\partial_{xx}\tilde{n}\big) \rangle\,dt.
\end{align}
The first term represents the amplitude of the state at time $T$. All the other terms are constraints enforced by Lagrange multipliers. The first of these fixes the initial amplitude while the remaining terms impose the dynamical constraints that $\bar{n}(x,t)$, $E(x,t)$, $\tilde{\phi}(x,t)$ and $\tilde{n}(x,t)$ all evolve in accordance with their respective nonlinear equations.
Maxima of this functional are found by seeking zeros of its first variational derivatives. We will refer to $\mathbf{X}(t)=(\bar{n},E,\tilde{\phi},\tilde{n})$ as the direct variables, and  $\boldsymbol{\chi}(t)=(\bar{N},F,\tilde{\Phi}^*,\tilde{N}^*)$ as the adjoint variables. For the variational derivatives to be zero, the following conditions must hold. For the direct variables
\begin{align}
0&=\partial_t\bar{n}-i\partial_x(\tilde{n}^*\tilde{\phi}-\tilde{n}\tilde{\phi}^*) \label{eqn:fwd1}\\
0&=\partial_tE-i\partial_x(\tilde{\phi}\partial_x\tilde{\phi}^*-\tilde{\phi}^*\partial_x\tilde{\phi})-\partial_{xx}E \\
0&=\partial_t\tilde{\phi}+i\tilde{\phi}(Sx+E) -i\tilde{n}-\partial_{xx}\tilde{\phi} \\
0&= \partial_t\tilde{n}+i\tilde{n}(Sx+E) -i\tilde{\phi}(\partial_x n -1) -\partial_{xx}\tilde{n}.\label{eqn:fwd4}
\end{align}
The adjoint variables must satisfy
\begin{align}
0&=\partial_t\bar{N}-\frac{i}{2}\partial_x(\tilde{N}^*\tilde{\Phi}-\tilde{N}\tilde{\Phi}^*) \label{eqn:adj1}\\
0&=\partial_tF-\frac{i}{2}(\tilde{n}\tilde{N}^*-\tilde{n}^*\tilde{N})
-\frac{i}{2}(\tilde{\phi}\tilde{\Phi}^*-\tilde{\phi}^*\tilde{\Phi})+\partial_{xx}F  \\
0&= \partial_t\tilde{\Phi}+i\tilde{\Phi}(Sx+E) -i\tilde{N}(\partial_x n -1) +2i\tilde{n}\partial_x\bar{N}
+2i\partial_x(\tilde\phi\partial_xF)+2i(\partial_x\tilde\phi)(\partial_xF)+\partial_{xx}\tilde{\Phi} \\
0&=\partial_t\tilde{N}+i\tilde{N}(Sx+E) -i\tilde{\Phi} -2i\tilde{\phi}\partial_x\bar{N}+\partial_{xx}\tilde{N}.\label{eqn:adj4}
\end{align}
The two sets of variables are linked through the optimality and compatibility conditions
\begin{align}
0 &=\boldsymbol{\chi}(0)-\lambda\mathbf{X}(0) \label{eqn:opt}\\
0 &=\boldsymbol{\chi}(T)-\mathbf{X}(T)  \label{eqn:comp}
\end{align}
(a full derivation of these equations is included in appendix \ref{app:derivAdj}).

These zeros are found via an iterative approach. An initial guess, $\mathbf{X}^0(0)$, is made. This can be evolved 
forwards in time to find $\mathbf{X}^0(T)$, guaranteeing equations (\ref{eqn:fwd1})-(\ref{eqn:fwd4}) are satisfied. We can then use this to `initialize' the adjoint variables $\boldsymbol{\chi}^0(T)=\mathbf{X}^0(T)$, satisfying equation (\ref{eqn:comp}). The adjoint variables can then be evolved backwards to satisfy conditions 
(\ref{eqn:adj1})-(\ref{eqn:adj4}) (note that the sign of the diffusive term indicates that this is the correct way
 to integrate these equations). At this point all the derivatives of $\mathscr{L}$ are zero except (\ref{eqn:opt}), 
 corresponding to $\partial\mathscr{L}/\partial\mathbf{X}(0)$, which  tells us how to update our initial guess
\begin{align}
\mathbf{X}^1(0)&=\mathbf{X}^0(0)-\epsilon\frac{\delta \mathscr{L}}{\delta \mathbf{X}(0)} \nonumber\\
&=\mathbf{X}^0(0)-\epsilon(\boldsymbol{\chi}^0(0)-\lambda\mathbf{X}^0(0)),
\end{align}
where $\lambda$ is chosen to maintain the initial state's amplitude across iterations. This procedure is then repeated 
until $\delta \mathscr{L}/\delta \mathbf{X}(0)$ is sufficiently small. 

\subsection{Linear transient growth}\label{sec:ltg}

Before calculating the minimal seed for this problem, we first take the exploratory step of considering the maximally growing disturbance of infinitesimal amplitude. 

The behavior of these is captured by the linearized equations
\begin{eqnarray}
\partial_t\tilde{\phi} &=& -i\tilde{\phi}Sx +i\tilde{n}+\partial_{xx}\tilde{\phi} \\ 
\partial_t\tilde{n} &=& -i\tilde{n}Sx -i\tilde{\phi}+\partial_{xx}\tilde{n} 
\end{eqnarray}
subject to the boundary conditions $\tilde{\phi}(L)=e^{-itSL}\tilde{\phi}(0)$ and 
$\tilde{n}(L)=e^{-itSL}\tilde{n}(0)$. The equations separate if we make the change of variables 
$a_\pm=\tilde{\phi}\pm i\tilde{n}$ leading to
\begin{equation}
\partial_t a_\pm = \pm\, a_\pm-ia_\pm Sx +\partial_{xx}a_\pm. \label{eqn:apm}
\end{equation}

Progress can be made analytically by making the ansatz 
\begin{equation}
a_+=A(t)\exp\{i(k-St)x\}, \, k=\frac{2n\pi}{L}.
\end{equation}
This ansatz arises more naturally when the 2D equations
are presented in the Lagrangian frame (moving with the fluid), where an initially plane wave solution does not change
wavenumber at late time, but the geometry secularly evolves with time (\citet{Johnson2005} denotes these
`shearing wave' modes {\it shwaves}).

The above ansatz leads to an equation for $A(t)$,
\begin{equation}
\frac{dA}{dt}+\big[(k-St)^2-1\big]A=0
\end{equation}
with solution 
\begin{equation}
A(t)=\exp\bigg\{(1-k^2)t+ kSt^2-\frac{1}{3}S^2t^3\bigg\}.
\end{equation}
Maxima and minima of this function can be found by seeking zeroes of the derivative of the exponent which occur when 
\begin{equation}
St-k=\pm 1.
\end{equation}
In order to achieve maximum growth we require a minimum at $t=0$ yielding the relationship $k_{lin}=1$,
leading to a maximum at $t_{lin}=\frac{2}{S}$. Putting this back into our equation for $A$ reveals the maximum 
growth being 
\begin{equation}
G_{lin}=\exp\bigg\{\frac{4}{3S}\bigg\}.
\end{equation}
A similar argument leads us to conclude $a_-$ must always monotonically decay, as is expected from inspection 
of equation (\ref{eqn:apm}).

\begin{figure}
\begin{center} 
\resizebox{0.9\textwidth}{!}{\includegraphics[angle=0]{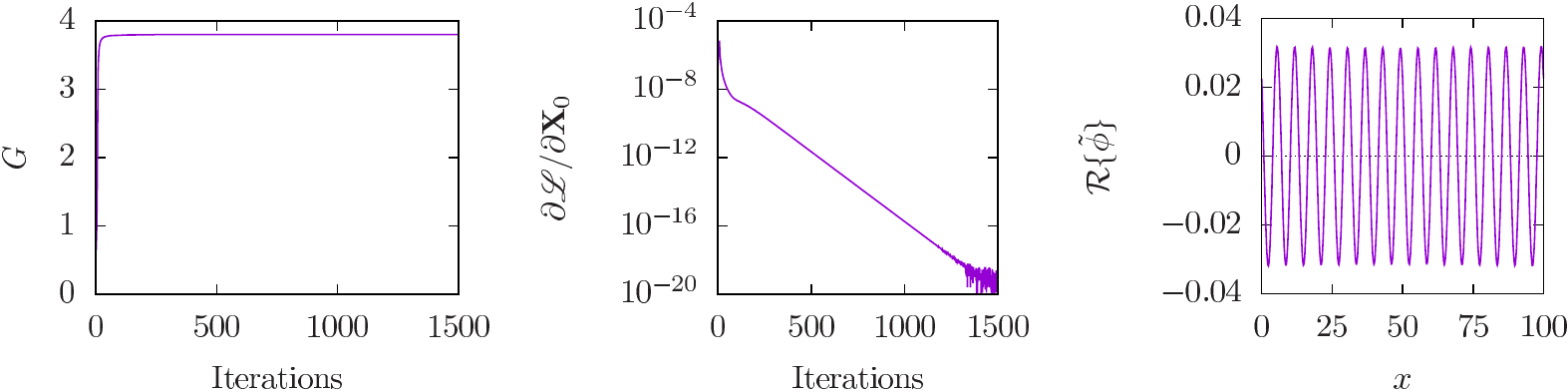}}
\end{center}
\caption{The linear energy growth optimal for $S=1.0$. From left to right: convergence of the growth 
rate as the algorithm iterates; decay of the amplitude of the gradient vector as the algorithm iterates; the 
real part of $\tilde{\phi}$ for the linear optimal.}
\label{fig:lin_conv}
\end{figure}

\begin{figure}
\begin{center} 
\resizebox{0.6\textwidth}{!}{\includegraphics[angle=0]{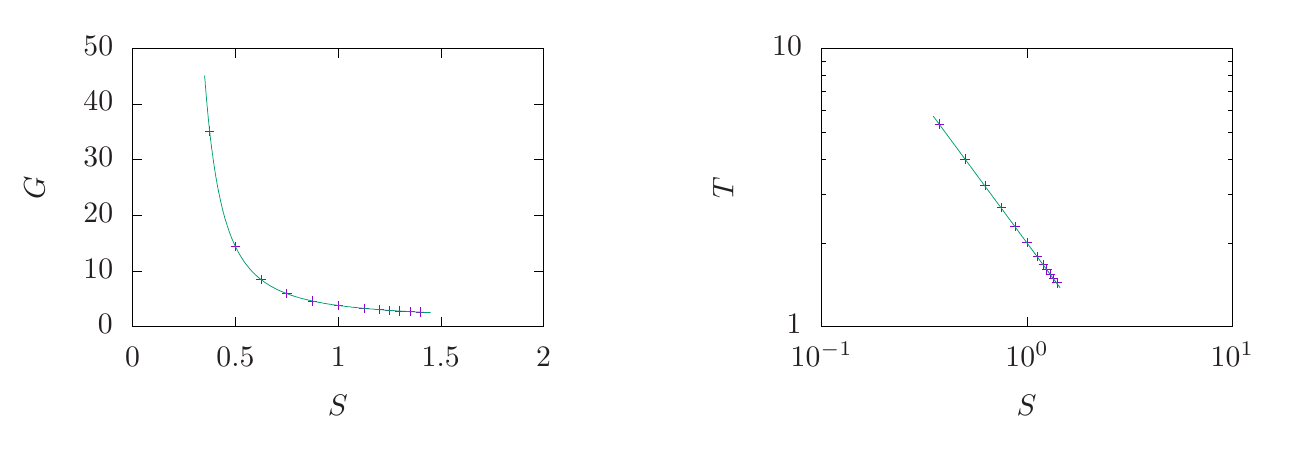}}
\end{center}
\caption{The amount of growth (left) and time taken for the growth to be achieved (right) by the linear optimal as 
a function of $S$. The lines are the predicted scalings of $T=2/S$ and $G=\exp\{4/3S\}$ and the crosses are the 
numerical results.}
\label{fig:g_t_lin}
\end{figure}

We confirm the linear result using the numerical method outlined previously. The procedure smoothly converges 
to the optimally growing disturbance (figure \ref{fig:lin_conv}). As expected this takes the form of a single Fourier mode in $\tilde{\phi}$ and $\tilde{n}$ with a phase shift of $\pi/2$ between them, and nothing in $\bar{n}$ or $E$. The optimal disturbance is independent of $S$, however the amount of growth and how long it takes does vary with the shear, as seen in figure \ref{fig:g_t_lin}. These scalings match with those derived above.

As an aside, this calculation also illustrates the linear instability of the system at $S=0$ as the amount of growth 
becomes infinite as long as $k^2<1$.

The linear transient dynamics of this model are similar to those in tokamak geometry at zero magnetic shear\cite{Highcock}. They are simpler, however, than the general tokamak case\cite{WaltzTrans,Bokshi2016}, where
the transient dynamics take the form of repeating periods of growth followed by damping (Floquet modes), as a geometrical coupling (magnetic shear) provides a means to couple modes of different radial wavenumber. 
In cases where the system is linearly stable, the maximum amplitude is reached after the first growth period, so the linear optimal would be sensitive only to behavior over this period; this may also be true for the nonlinear problem.

One approach for characterizing linear transient growth for tokamak instabilities is described in \citet{Roach09}. An
effective growth rate is defined by $log(A)/t$ where $t$ is the time taken for the mode to grow by some predetermined
factor $A$. This can be applied to any initial condition, even if it eventually decays. The method here can also determine
such effective growth rates, and choosing to optimize the initial state uniquely defines it.

\subsection{Nonlinear optimals}

We now turn our attention to finding the minimal seed by calculating optimally growing disturbances with finite amplitude. In theory this is a simple extension of the linear problem, but in practice it suffers from a number of technical problems that need to be understood.

The first is that although the adjoint equations are in fact linear in the adjoint variables, they do require that the 
forward variables are stored for all times as they also appear in the equations. For large systems (or long time runs) this can become memory intensive. This can be worked around by introducing checkpointing where the forward state is only saved a smaller number of discrete times (say $\tau_1$, $\tau_2$, $\dots ,\tau_N$). During the backwards run the forward variables can be reconstructed in a piecemeal fashion with each section of the direct variables recovered when it is needed by integrating forward again. 

Choosing the target time is also a challenge. For the linear problem we optimized for $T$ to maximize the growth. 
Although that can also be done here,\cite{rabin12} it is easier -- and more consistent with our argument -- to simply fix the target time as something large. Although it is tempting to set $T$ to be as large as can be feasibly managed, this leads to complications and in particular convergence can be very slow (in terms of the number of iterations required) and perhaps not even possible. This is because as most states will evolve towards zero after an initial transient, $\mathbf{x}_T$ can be susceptible to numerical noise and the gradient vector becomes a poor approximation of the true one. For our purposes we settled on taking $T=3T_{lin}$ which is large enough for the dynamics to develop but not for damaging levels of viscous decay to occur.

Iterative approaches are not guaranteed to converge. For small initial amplitudes, convergence should be straight forward and the result is expected to be the same as the linear one (or at most a small modification to it). Once larger amplitudes are encountered, however, it seems less clear what to expect. In general, if the initial states being considered evolve into a chaotic regime then we don't expect convergence at all. Small changes to the initial state lead to large changes in the energy growth and the problem effectively becomes non-smooth. If we consider a system for which the edge separates chaos from smooth deterministic behavior, then we expect subcritical amplitudes (below the edge) runs to converge and supercritical amplitudes (above the edge) to fail to converge. If the edge itself is chaotic then critical amplitude runs will also not converge. 

This gives a simple criterion for identifying the critical amplitude by running the algorithm at a sequence of amplitudes and identifying where convergence ceases to be possible. This abrupt cut off becomes smoothed when a finite optimization time is taken. As the critical amplitude is approached, chaos takes longer and longer to develop until this time frame 
is larger than the chosen optimisation time. The approach is further hindered when either the non-trivial state or the edge is not chaotic. We have already seen the latter is the case for the PI model, and for some larger values of $S$ the former is also true. If the non-trivial state is deterministic then convergence is possible above the edge as smoothness is not lost. This is not a problem as the amount of growth should be distinct from that seen below or within the edge. If the edge is deterministic (i.e. there is a simple traveling wave within the edge as here) then for finite times disturbances close to, but on either side of, the edge will both evolve to something approximating the edge state. Although given enough time these would diverge away from the edge state, if they are arbitrarily close to the edge then the amount of time required is arbitrarily large. The amount of growth on either side of the edge after more moderate times will be comparable and the only way to determine whether the edge has been crossed is to use the optimal as the initial condition for a longer time simulation.

\subsection{Critical amplitudes}

\begin{figure}
\begin{center} 
\resizebox{0.9\textwidth}{!}{\includegraphics[angle=0]{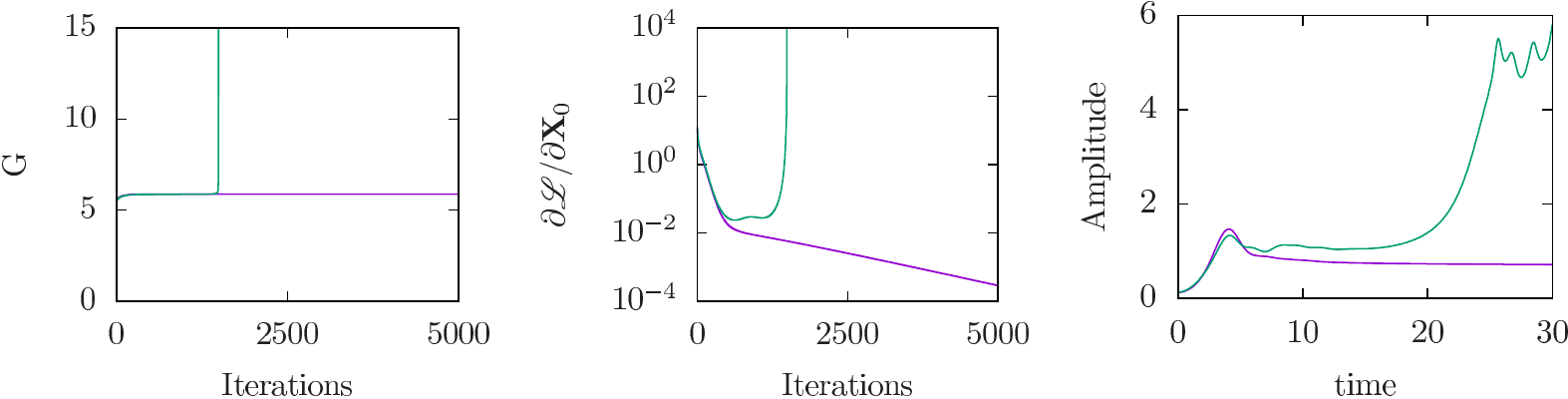}}
\end{center}
\caption{The nonlinear energy growth optimal at $S=0.5$. From left to right: convergence of the growth rate 
as the algorithm 
iterates; decay of the amplitude of the gradient vector as the algorithm iterates; the amplitude evolution of the 
nonlinear optimal and the first turbulent seed found by the algorithm. In all cases the purple line corresponds to 
$A=0.127$ and the green line corresponds to $A=0.129$.}
\label{fig:nltg_conv_0.5}
\end{figure}

In figure \ref{fig:nltg_conv_0.5} we show the results of the algorithm for $S=0.5$ for two different 
initial amplitudes -- $A=0.127$ and $0.129$. Although the amplitudes are very similar, fundamentally different 
results are obtained. The lower 
amplitude smoothly converges. The larger amplitude starts to follow the same path before diverging. By checking 
how each of the states the algorithm iterates through evolve in time, we see that this divergence corresponds to 
the point at which we begin to see turbulence in the system. The last few states that are iterated through at 
$A=0.129$ all lead to turbulent episodes and so are turbulent seeds above the edge. All of the states found for $A=0.127$ 
are below the edge, including the final converged optimal. We conclude that the critical amplitude, $A_C$, is sandwiched in between these two values. In this way we bound the minimum amplitude required to trigger turbulence.

\begin{figure}
\begin{center} 
\resizebox{0.6\textwidth}{!}{\includegraphics[angle=0]{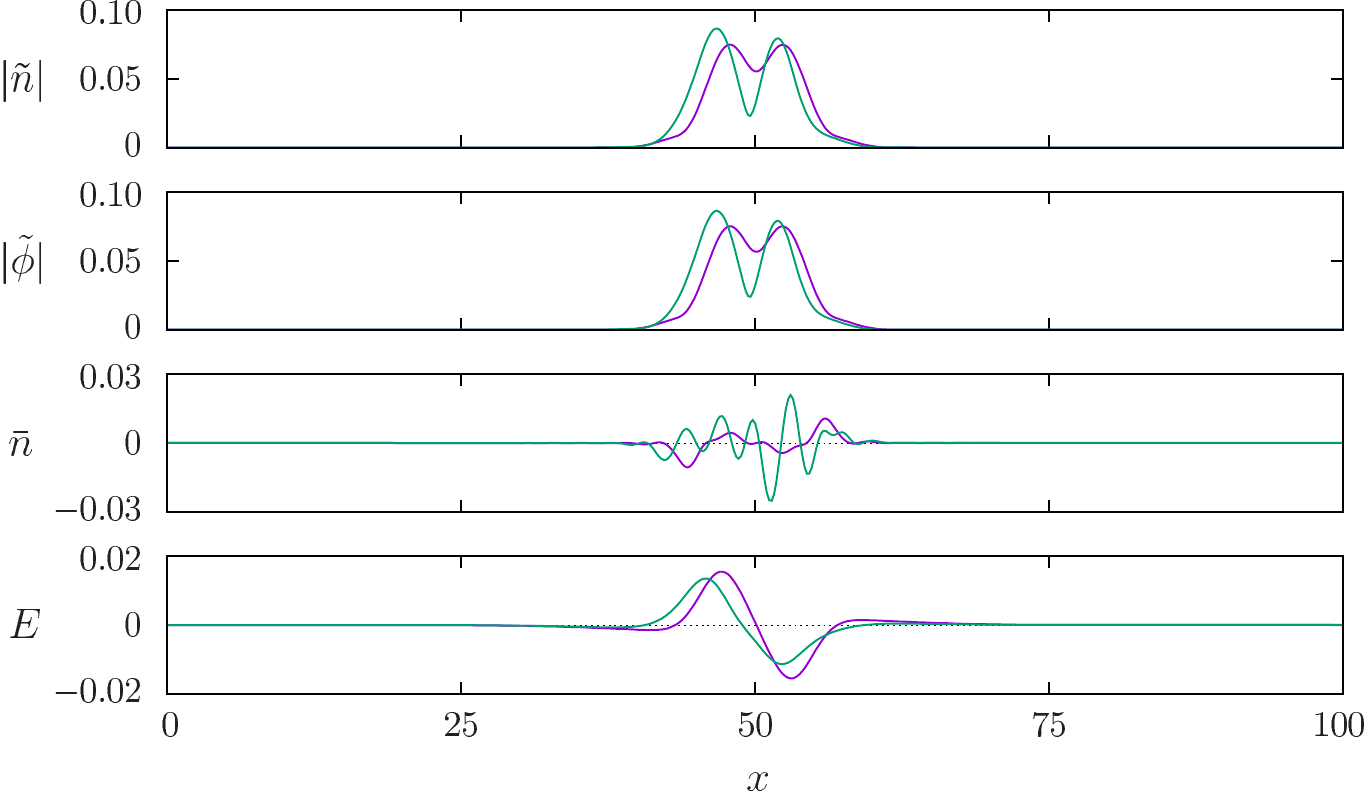}}
\end{center}
\caption{The nonlinear energy growth optimal ($A=0.127$, purple) and the turbulent seed ($A=0.129$, green) at $S=0.5$. The two states are both strongly localized and are similar in form, the principal difference being in the $\bar{n}$ fields.}
\label{fig:minseed_0.5}
\end{figure}

\begin{figure}
\begin{center} 
\resizebox{0.45\textwidth}{!}{\includegraphics[angle=0]{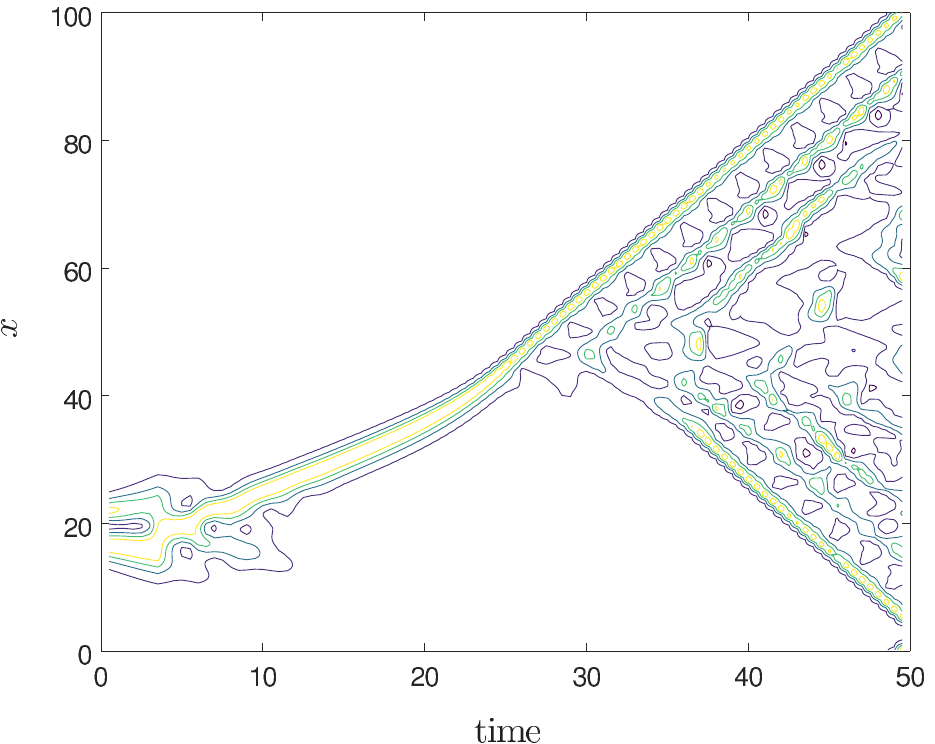}}
\end{center}
\caption{Evolution of the first turbulent seed for $S=0.5$. The seed is localized and its initial evolution reflects 
this as it approaches the localized traveling wave solution within the edge. From here it then evolves into higher energy 
turbulence which spreads to fill the domain. The contours are based upon the flux, 
$Q=\tilde{n}^*\tilde{\phi}-\tilde{n}\tilde{\phi}^*$, but at each time it is normalized by the instantaneous maximum 
value of the flux at that time: $Q'=Q(x,t)/\max_x{Q(x,t)}$. This normalization is chosen as the turbulent seed grows by 
several orders of magnitude during it evolution. Contours are $\pm 5\%$, $\pm 25\%$, $\pm 50\%$ and 
$\pm 75\%$ of the instantaneous maximum flux.}
\label{fig:minseed_q}
\end{figure}

Both the nonlinear optimal and the first turbulent seed (the first initial condition found by the algorithm that initiates turbulence) are localized and show similar structure (figure \ref{fig:minseed_0.5}). The localization is to be expected. This allows the disturbance to have local amplitudes large enough to nonlinearly self-interact (a 
necessity for turbulence) while simultaneously keeping the total amplitude small. Despite the minimum seed being localized in space, it still initiates space-filling turbulence (figure \ref{fig:minseed_q}). After an initial transient ($0\leq t\lesssim 10$),the flow approaches the traveling wave solution ($10\lesssim t\lesssim 25$) before growing into the fully turbulent state ($t\gtrsim 25$). This turbulence spreads in both directions to fill the entire domain. The amount of time 
this takes is a function of how close to the critical amplitude the initial amplitude is. The initial 
growth towards traveling wave takes an amount of time approximated by the time-scale of the linear transient growth. As it 
approaches the traveling wave, how close it comes to that solution (and hence how long the next stages take) depend on 
the initial amplitude as well as the leading eigenvalues of the traveling wave solution. 

\begin{figure}
\begin{center} 
\resizebox{0.9\textwidth}{!}{\includegraphics[angle=0]{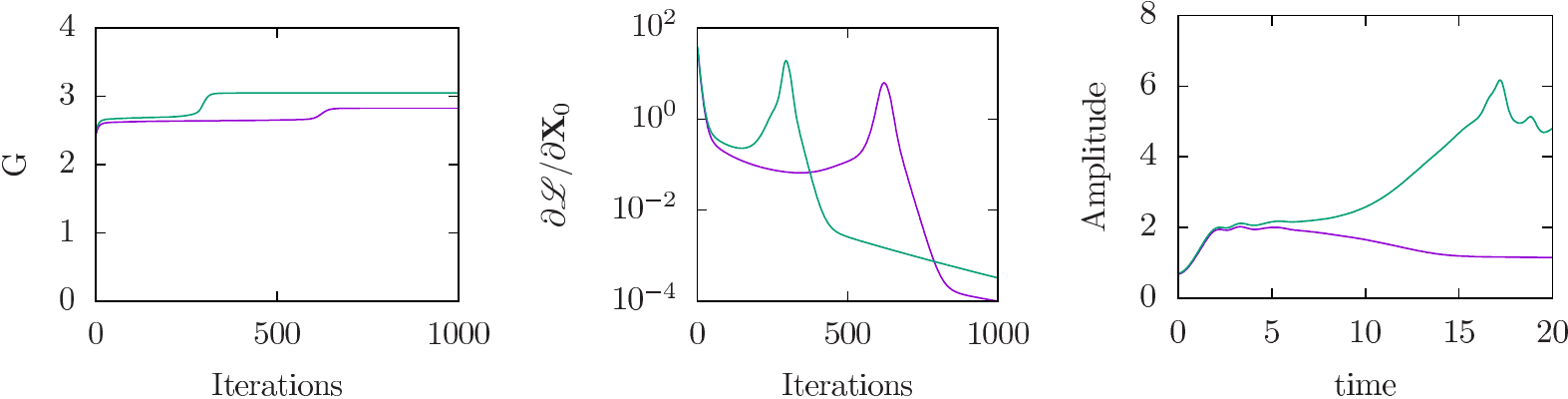}}
\end{center}
\caption{The nonlinear energy growth optimal at $S=1.0$. From left to right: convergence of the growth rate 
as the algorithm 
iterates; decay of the amplitude of the gradient vector as the algorithm iterates; the amplitude evolution of the 
nonlinear optimal and the first turbulent seed found by the algorithm. In all cases the purple line corresponds to 
$A=0.69$ and the green line corresponds to $A=0.71$.}
\label{fig:nltg_conv_1.0}
\end{figure}

The picture changes a little as we move to higher values of the background shear. At $S=1.0$, the observed turbulence 
is much smoother and more slowly varying. This is also true of the behavior within the edge. Because of this, 
initial conditions near the edge spend much longer near to the edge and our iterative approach converges 
above the edge even for these moderate choices of $T$ (figure \ref{fig:nltg_conv_1.0}). Due to this care must 
be taken to identify the point at which the edge is crossed by examining individual runs rather than simply finding 
the point at which convergence is no longer possible.

\begin{figure}
\begin{center} 
\resizebox{0.6\textwidth}{!}{\includegraphics[angle=0]{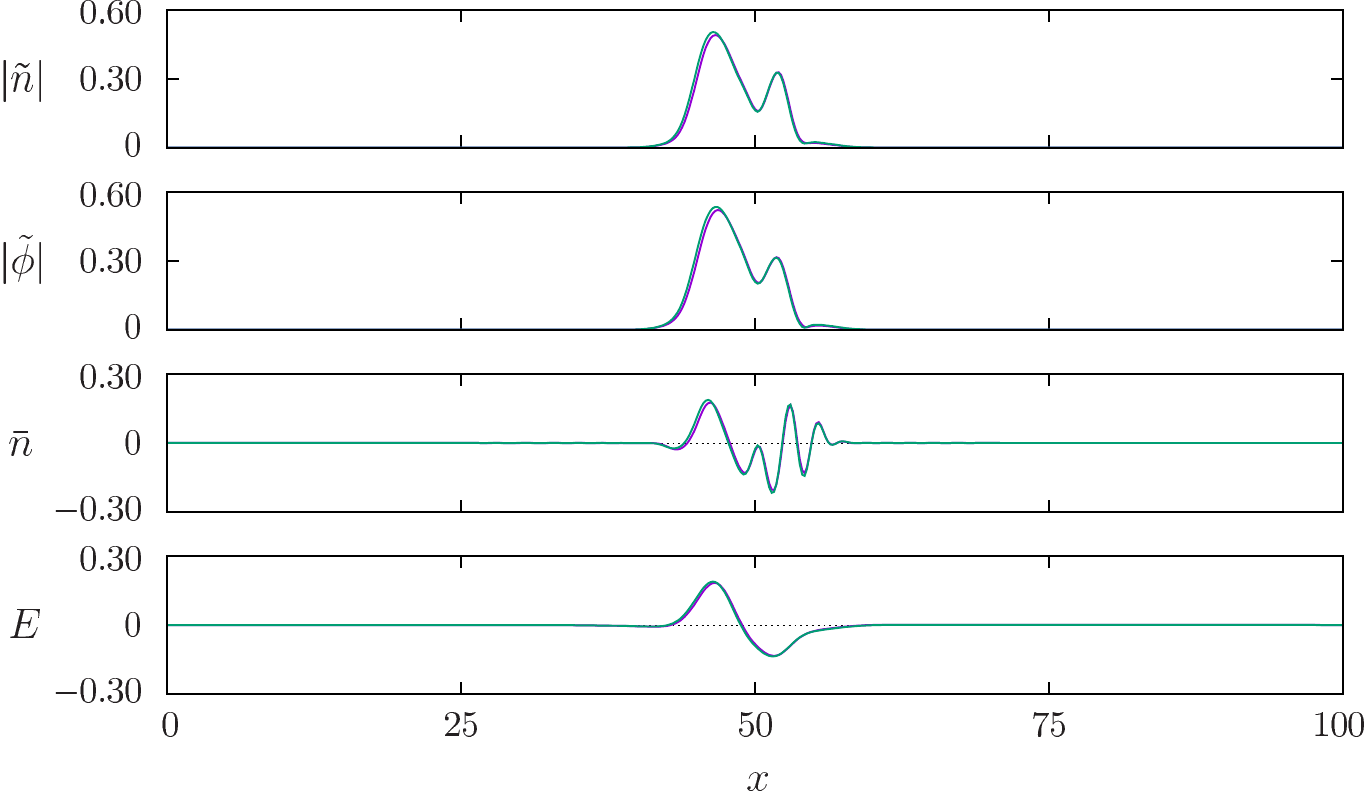}}
\end{center}
\caption{The nonlinear energy growth optimal ($A=0.69$, purple) and the turbulent seed ($A=0.71$, green) at $S=1.0$. 
The two states are barely distinguishable and are both strongly localized.}
\label{fig:minseed_1.0}
\end{figure}

Because we are able to converge so close to the edge, we are able to much more accurately pin down the minimal seed. 
The fact the energy growth optimals on either side of the edge are so similar in form (figure \ref{fig:minseed_1.0}) strongly indicates that the minimal seed, which is sandwiched between them, must also be very similar.

\begin{figure}
\begin{center} 
\resizebox{0.5\textwidth}{!}{\includegraphics[angle=0]{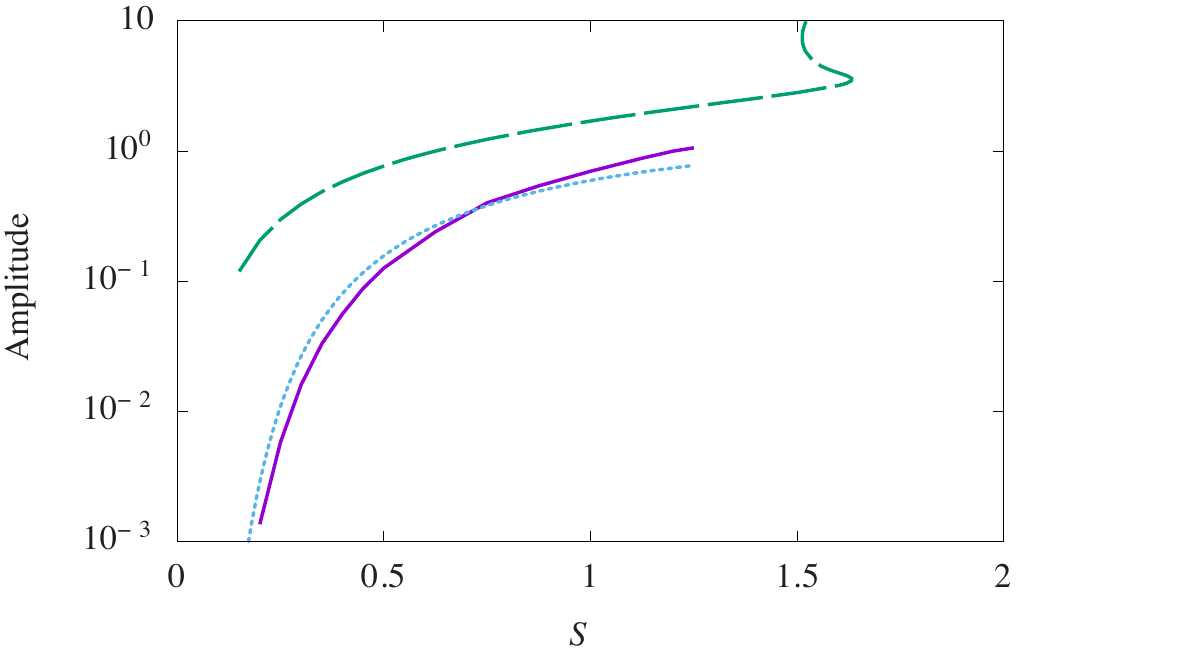}}
\end{center}
\caption{The amplitudes of the minimum seed (solid) and of the traveling wave solutions (long dashes) as functions of the background shear. Also included is the predicted critical amplitude from the semi-analytical analysis (short dashes) found in section \ref{sec:approx}. Unsurprisingly the minimal seed triggers turbulence at amplitudes significantly below the edge state. The semi-analytic approach agrees well with the minimal seed, especially for small values of $S$ where the approach is better justified.}
\label{fig:amp_E}
\end{figure}

By repeating this approach for a range of values of $S$, we are able to see how the minimum amplitude of the edge scales with the background shear. In figure \ref{fig:amp_E} we plot the amplitudes of the minimum seed and the traveling solution as functions of $S$. We are unable to continue the minimum seed calculation to values of $S$ quite as high as the traveling wave goes because turbulence starts to become difficult to sustain here. The method adopted can only be expected to work when turbulence is sustained (otherwise after long times all disturbances must return to the laminar) and there is a clear differentiation between the amplitude of the observed turbulence and that of the edge. If these amplitudes become too similar then the optimization algorithm has no reason to find seeds that lead to turbulence.

One might attempt to trigger turbulence with either the linear optimal or the most unstable linear mode, but this would fail as both these disturbances are plane waves which must eventually decay, even if they transiently attain amplitudes much larger than typical turbulence levels (although small amounts of noise in the initial perturbation may be enough to allow a transition). This is an indication of the potential for an approach based simply on transient linear growth to be misleading in, for example, identifying the most dangerous perturbations. 

Note that the nonlinear stability of plane wave solutions is a generic feature of periodic shearing-box systems with convective nonlinearities\cite{Johnson2005,Squire2014}. Even for systems for which a plane-wave linear optimal can trigger turbulence, the critical amplitude will scale with the system size, unlike that of the nonlinear optimal, which usually becomes independent of system size. Only the nonlinear analysis captures the fact that turbulence triggering is a local process.

\section{Semi-analytic approximation}\label{sec:approx}

The minimal seed calculation performed in the previous section provides a means to calculate the critical amplitude but does not directly explain which mechanisms are responsible for allowing a small initial perturbation to evolve to a nonlinearly saturated state. In this section we present a semi-analytic approach to approximating the minimal seed, and hence provide a simple closed form estimate for the critical amplitude,
thus developing some insight into the processes occurring during the initial transient.

The approach revolves around the assumption that for turbulence to be sustained, the background shear needs to be ameliorated (this is justified partly by examining the time evolution of $E$ in earlier results).
This occurs when $\partial E/\partial x$ is of a similar size to $S$. This is exactly the point where the linear and nonlinear terms in the time evolution of $\tilde{\phi}$ and $\tilde{n}$ become comparable. We assume, for the sake of obtaining a rough estimate, that up to this point we can use the linear time evolution for $\tilde{\phi}$ and $\tilde{n}$ as in section \ref{sec:ltg}.

We consider the evolution of an initial disturbance, $\tilde{\phi}(x,0)$. The wave density is then given by 
$\tilde{n}=-i\tilde{\phi}$ (in our previous notation this gives $a_-=0$), while the remaining two components, 
$E$ and $\bar{n}$ are taken to be zero. After applying the Fourier transform 
\begin{equation}
\tilde{\Phi}(k,t)=\int_{-\infty}^{\infty}\tilde{\phi}(x,t)e^{ikx}\,dx
\end{equation}
we know this new field evolves as 
\begin{equation}
\tilde{\Phi}(k,t)=\tilde{\Phi}(k,0)\exp \bigg\{(1-k^2)t+ kSt^2-\frac13S^2t^3\bigg\}.
\end{equation}
As the disturbance evolves, its effective wave number changes linearly with time. This can be scaled out by 
introducing the new variable $\tilde{k}=k+St$ in which case
\begin{equation}
\tilde{\Phi}(\tilde k,t)=\tilde\Phi(\tilde k-St,0)\exp \bigg\{t -\frac{1}{3S}\big[ (\tilde k +St)^3 
- \tilde k^3\big]\bigg\}. \label{eqn:Phi}
\end{equation}

We are interested in the evolution of $E$, governed by
\begin{equation}
\partial_tE=i\partial_x(\tilde{\phi}\partial_x\tilde{\phi}^*-\tilde{\phi}^*\partial_x\tilde{\phi})+\partial_{xx}E.
\label{eqn:E}
\end{equation}
This will be dominated by the behavior when $\tilde\phi$ is at its largest. As previously seen, the amplification factor 
in equation (\ref{eqn:Phi}) has a maximum at $t=t_{lin}=2/S$, $\tilde k=\tilde k_{lin}=-1$. Close to this we 
can approximate the behavior of $\tilde\Phi$ by
\begin{equation}
\Phi(k,t)=\Phi(k-St,0)\exp \bigg\{\frac{4}{3S} -
\frac{1}{S} \big[k-k_{lin} +S(t-t_{lin})\big]^2 - \frac{1}{S}\big[ k-k_{lin}\big]^2\bigg\}
\end{equation}
plus higher order terms which we have dropped, along with the $\sim$'s, for clarity.

We make the ansatz
\begin{equation}
\phi(x,0)=Ae^{-\sigma^2x^2}e^{ik_{lin}x} \Rightarrow \Phi(k,0)=\frac{A\sqrt{\pi}}{\sigma}e^{-(k-k_{lin})^2/4\sigma^2},
\end{equation}
corresponding to the plane wave linear optimal but modulated by a Gaussian and so localized in space. 
The choice of $\sigma^2=S/4$ is made here, both because it simplifies the algebra, and also because
this gives the most localization of the initial condition in real space while still exciting mostly
modes with near-maximal linear amplification. With this choice the behavior close to its maximum amplitude is
\begin{equation}
\Phi(k,t)=2A\sqrt{\frac{\pi}{S}}\exp\bigg\{\frac{4}{3S}\bigg\}
\exp\bigg\{-\frac{3}{S}(k-k_{lin})^2-2S(t-t_{lin})^2\bigg\},
\end{equation}
which in real space gives
\begin{equation}
\phi(x,t)=\frac{A}{\sqrt{3}}
e^{\frac{4}{3S}-2S(t-t_{lin})^2}e^{-Sx^2/12}e^{ik_{lin}x}.
\end{equation}

Inserting this into equation (\ref{eqn:E}) we find
\begin{equation}
\frac{\partial E}{\partial t} = \frac29 A^2Sk_{lin} x e^{-Sx^2/6}e^{\frac{8}{3S}-4S(t-t_{lin})^2} + \frac{\partial^2 E}{\partial x^2}.
\end{equation}
We estimate the maximum value that $\partial E /\partial x$ achieves by neglecting the diffusive term on the RHS and first integrating with 
respect to $t$. Setting $t=t_{lin}$ will give us the most shear, which is in turn evaluated by differentiating with 
respect to $x$ to get
\begin{equation}
\max_{x,t} \frac{\partial E}{\partial x} = -\frac19 A^2 \sqrt{\pi S}\exp\bigg\{\frac{8}{3S}\bigg\}.
\end{equation}
For a broader wavepacket with the same norm, the $x$ derivatives would have lead to a smaller prefactor;
an explicit optimization of $\sigma$ would balance this against better localization in wavenumber space for maximum transient growth.
$\partial E/\partial x$ balances the background shear, $S$, when
\begin{equation}
A^2= 9 S^{1/2}\pi^{-1/2}\exp\bigg\{-\frac{8}{3S}\bigg\}.
\end{equation}

In figure (\ref{fig:amp_E}) we include this predicted amplitude for transition. The agreement between this prediction 
and the amplitude of the minimum seed is good, especially at low $S$ which fits with our assumptions. As a means of
estimating the critical amplitude, this approach captures several key aspects of the dynamics. Firstly, the scaling
of the true critical amplitude seems to be dominated by the transient growth factor, indicating that linear transient
growth is the main mechanism able to amplify the initial seed perturbations. Secondly, the spatial localization of the initial condition (the full-width at half-maximum of $|\tilde \phi^2|$)  and transition state is well predicted (to within a factor of 2). Thirdly, the peak initial wavenumber of the minimal seed is similar to that of the semi-analytical initialization.

It has been shown that there is a correlation between the region where turbulence may be sustained and and the level of linear transient amplification for some tokamak cases\cite{VanWyk2017}, and this kind of semianalytical treatment might be able characterise the role of nonlinear processes in the transition from turbulence to the laminar regime (other approaches are also promising\cite{Connaughton2015}). Note, however, that the properties of the transition to turbulence and those of the saturated turbulent state are not directly related: for example, the amplitude where the nonlinear transition occurs increases with $S$, whereas the average turbulence amplitude decreases with $S$.

The major benefit of this semi-analytical approach is that it allows a quantitative understanding of the mechanisms allowing small perturbations to evolve towards a nonlinearly self-sustaining state, and provides scaling estimates. The full minimal seed calculation is a necessary preliminary step, from which the active mechanisms can be determined. The assumption, for example, that the initial evolution of $E$ occurs due to the shape of the amplified wave-packet was based on inspecting the time-evolution of the minimal seed: this is rather different to a common assumption elsewhere that the modulational instability is key to zonal-flow growth during nonlinear saturation\cite{Diamond_review}.

\section{Conclusions and discussions}\label{sec:conc}

We have presented two methods useful in the analysis of the transition to turbulence in linearly stable systems. Such subcritical configurations are exhibited by many plasma systems of interest, e.g. the suppression of temperature-gradient driven drift instabilities by poloidal zonal flows in tokamaks. For these subcritical configurations, linear approaches struggle to give meaningful results, while the nonlinear approaches presented here offer fresh insight.

The edge of chaos is the boundary between the basins of attraction of the laminar state and the turbulent one. 
By tracking trajectories confined to this boundary, we found that it is dominated by a traveling wave solution 
which is an attractor within the edge. This traveling wave is localized in space, even for shear rates at 
which turbulence is space filling. This is indicative of fact that one can initiate global turbulence with local
disturbances. Similar behavior is seen in classical shear flows\cite{duguet09,avila13}. By tracking the traveling 
wave through phase space we are able to quickly track the typical amplitude of the edge. Further, the maximum shear rate for which turbulence is sustained seems to correspond to the the maximum shear rate at which the traveling wave exists. It should be noted that the traveling wave ceasing to exist does not have to exactly correspond to the point at which turbulence ceases to exist,\cite{avilaSci11} but it is expected to be related. At this point, something fundamental has to change in the nature of the edge for this system and hence for the system as a whole.

If the edge state represents the \emph{typical} turbulence inducing disturbance, the minimal seed is the most 
\emph{dangerous} disturbance --  the smallest disturbance capable of triggering turbulence. As with the edge state, this seed is localized. This should be expected as it allows the disturbance to be of low total amplitude while still being locally large enough to exhibit nonlinear behavior. 

The play off between local and global amplitude is explored in the semi-analytic approach we put forward which qualitatively captures the key dynamics. The semi-analytic approximation is shown to offer a good agreement with the nonlinearly computed minimal seed amplitude. It may offer a computationally efficient approach in systems where performing a full nonlinear optimization is not feasible (an important consideration in kinetic models). However, the principal outcome is a quantitative understanding of the mechanisms that allows small perturbations to evolve towards a nonlinearly self-sustaining state. 

\begin{figure}
\begin{center} 
\resizebox{0.5\textwidth}{!}{\includegraphics[angle=0]{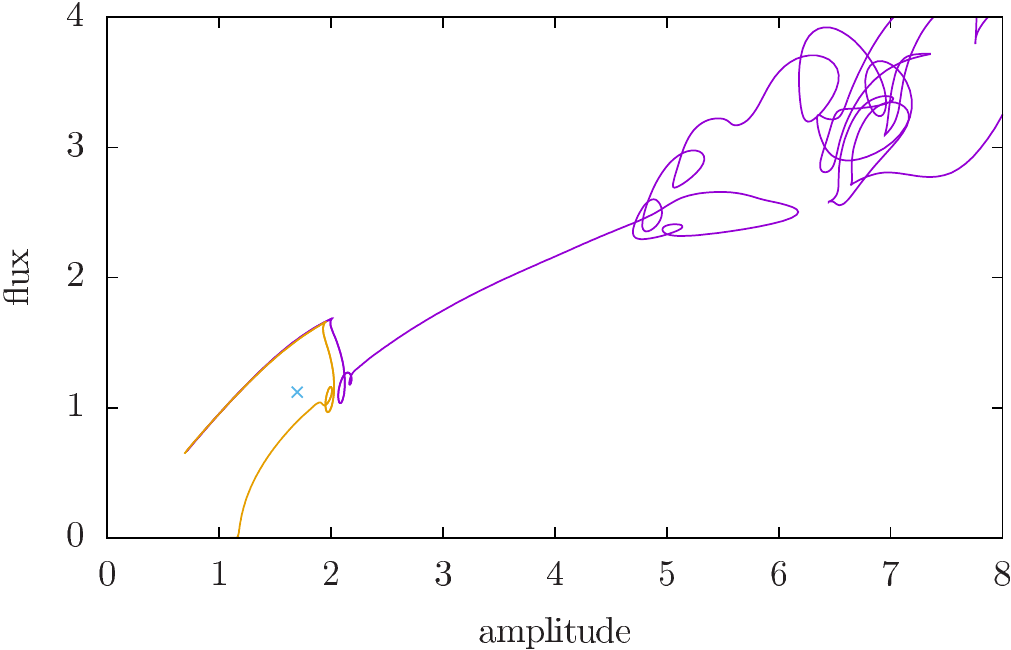}}
\end{center}
\caption{Phase space plot of flux ($Q=\int_0^L \tilde{n}^*\tilde{\phi}-\tilde{n}\tilde{\phi}^*\,dx$) against amplitude for the transition to turbulence at $S=1.0$. The orange line is the nonlinear energy growth optimal for $A=0.69$ while the purple line is the optimal for $A=0.71$, shown in figure \ref{fig:minseed_1.0}. Both initially evolve in a very similar manner, approaching the traveling wave for this shear, marked `$\times$'. As they approach the edge state the two trajectories diverge to the laminar state and turbulent state respectively.}
\label{fig:phase}
\end{figure}

The two approaches (edge tracking and minimal seed calculation) link to each other. In figure \ref{fig:phase} 
we plot the evolution in phase space of the two states sandwiching the minimal seed at $S=1.0$, previously shown in figure \ref{fig:minseed_1.0}. As expected both of them initially display closely matching evolutions as they track their way along the edge. Eventually they start to diverge, one relaminarising while the other exhibits turbulent behavior. The trajectories depart one another as they reach the traveling wave solution, marked with a cross.
 
The future task is to apply these methods to the full plasma problem. Edge tracking is a straight forward tool that 
can be readily implemented within already existing codes. Although it seems unlikely to lead to the discovery of 
attracting solutions embedded within the edge, unstable solutions may be identified. Close approaches to these
exact solutions within the edge can be found by seeking moments of near-recurrence.\cite{dwk08} Newton-Krylov methods based on well-established routines such as GMRES\cite{gmres} reduce the computational intensity of solution tracking. The probability of successful convergence can be enhanced by using so-called globally convergent methods.\cite{dennisSchnabel} Calculating minimal seeds for large problems is also computationally intensive. A possible way to reduce the overhead would be to instead consider smaller domains in which turbulence may even be space filling. In the case of shear flows the minimal seeds for such reduced problems appear to closely mirror those of the full, non-periodic problem.\cite{pwk15} For the PI model considered, the difference in amplitudes between the edge state and the minimal seed is not excessive. This is somewhat  atypical\cite{PK10,rabin12,duguet13} and it seems likely that for the full plasma problem this energy gap will increase drastically. 

Despite these computational hurdles, the methods presented here seem ripe for applying to tokamak problems.\cite{mptPRL} The limitations of linear approaches are becoming increasingly clear and so fully nonlinear methods allow for a 
fresh take. Ultimately, stabilizing subcritical plasma flow requires a means of quantifying the \emph{nonlinear} 
stability of a system. Both edge tracking and minimal seeds offer a meaningful way of doing so. Endeavors 
to incorporate these into classical control have already shown promise.\cite{kawahara05,rabin14}

\section{Acknowledgments}\label{sec:ack}
We thank Zaid Iqbal for useful discussions. 
C. C. T. Pringle is partially supported by EPSRC grant no. EP/P021352/1.
B. McMillan is partially supported by EPSRC grant no. EP/N035178/1.
B. Teaca is partially supported by EPSRC grant No. EP/P02064X/1.

\appendix
\section{Derivation of the adjoint equations}\label{app:derivAdj}
In section \ref{sec:nltg} we defined the functional
\begin{align}
\mathscr{L}=&\langle \bar{n}(x,T)^2+E(x,T)^2+\tilde{\phi}(x,T)^2+\tilde{n}(x,T)^2 \rangle
 -\lambda\big(\langle \bar{n}(x,0)^2+E(x,0)^2+\tilde{\phi}(x,0)^2+\tilde{n}(x,0)^2 \rangle -A_0\big)\nonumber \\
&-\int_0^T \langle \bar{N}.\big(\partial_t\bar{n}-i\partial_x(\tilde{n}^*\tilde{\phi}-\tilde{n}\tilde{\phi}^*)\big) \rangle\,dt 
 -\int_0^T \langle F.\big(\partial_tE-i\partial_x(\tilde{\phi}\partial_x\tilde{\phi}^*-\tilde{\phi}^*\partial_x\tilde{\phi})-\partial_{xx}E\big) \rangle\,dt \nonumber \\
&-\int_0^T\langle \tilde{\Phi}^*.\big(\partial_t\tilde{\phi}+i\tilde{\phi}(Sx+E) -i\tilde{n}-\partial_{xx}\tilde{\phi}\big) \rangle\,dt
-\int_0^T\langle \tilde{N}^*.\big(\partial_t\tilde{n}+i\tilde{n}(Sx+E) -i\tilde{\phi}(\partial_x \bar{n} -1) -\partial_{xx}\tilde{n}\big) \rangle\,dt
\end{align}
which we sought to maximize over all possible choices of initial condition. In order to find these optimals, 
we used the variational derivatives found here.

\begin{align}
\delta \mathscr{L}=&\langle \delta\bar{n}(x,T).\bar{n}(x,T)+\delta E(x,T).E(x,T)+
\delta\tilde{\phi}(x,T).\tilde{\phi}(x,T)+\delta\tilde{n}(x,T).\tilde{n}(x,T) \rangle \nonumber \\ 
&-\delta \lambda.\big(\langle \bar{n}(x,0)^2+E(x,0)^2+\tilde{\phi}(x,0)^2+\tilde{n}(x,0)^2 \rangle -A_0\big)\nonumber \\
&-\lambda .\langle \delta\bar{n}(x,0).\bar{n}(x,0)+\delta E(x,0).E(x,0)+
\delta\tilde{\phi}(x,0).\tilde{\phi}(x,0)+\delta\tilde{n}(x,0).\tilde{n}(x,0) \rangle
\nonumber \\
&-\int_0^T \langle \delta\bar{N}.\big(\partial_t\bar{n}-i\partial_x(\tilde{n}^*\tilde{\phi}-\tilde{n}\tilde{\phi}^*)\big) \rangle\,dt
-\int_0^T \langle \bar{N}.\delta\big(\partial_t\bar{n}-i\partial_x(\tilde{n}^*\tilde{\phi}-\tilde{n}\tilde{\phi}^*)\big) \rangle\,dt\nonumber \\
&-\int_0^T \langle \delta F.\big(\partial_tE-i\partial_x(\tilde{\phi}\partial_x\tilde{\phi}^*-\tilde{\phi}^*\partial_x\tilde{\phi})-\partial_{xx}E\big) \rangle\,dt
-\int_0^T \langle F.\delta \big(\partial_tE-i\partial_x(\tilde{\phi}\partial_x\tilde{\phi}^*-\tilde{\phi}^*\partial_x\tilde{\phi})-\partial_{xx}E\big) \rangle\,dt \nonumber \\
&-\int_0^T\langle \delta\tilde{\Phi}^*.\big(\partial_t\tilde{\phi}+i\tilde{\phi}(Sx+E) -i\tilde{n}-\partial_{xx}\tilde{\phi}\big) \rangle\,dt
-\int_0^T\langle \tilde{\Phi}^*.\delta\big(\partial_t\tilde{\phi}+i\tilde{\phi}(Sx+E) -i\tilde{n}-\partial_{xx}\tilde{\phi}\big) \rangle\,dt\nonumber \\
&-\int_0^T\langle \delta\tilde{N}^*.\big(\partial_t\tilde{n}+i\tilde{n}(Sx+E) -i\tilde{\phi}(\partial_x \bar{n} -1) -\partial_{xx}\tilde{n}\big) \rangle\,dt
-\int_0^T\langle \tilde{N}^*.\delta\big(\partial_t\tilde{n}+i\tilde{n}(Sx+E) -i\tilde{\phi}(\partial_x \bar{n} -1) -\partial_{xx}\tilde{n}\big) \rangle\,dt.
\end{align}
The majority of these terms are in an appropriate from, but the terms in the form 
\begin{equation*}
\int_0^T \langle \chi.\delta\big(\partial_t \textrm{x} - F(\textrm{x})\big)\rangle \,dt
\end{equation*}
require manipulation. The time-derivative terms are all similar and in general can be dealt with as
\begin{align}
\int_0^T\langle A.\delta(\partial_t B)\rangle \,dt = & \int_0^T\langle \partial_t(A.\delta B) -\delta B.\partial_t A\rangle \,dt \nonumber \\
=&\langle\delta B(T). A(T)\rangle -\langle \delta B(0).A(0)\rangle -\int_0^T \langle \delta B.\partial_t A\rangle \,dt.
\end{align}
The diffusive term yields
\begin{align}
\int_0^T\langle F.\delta(\partial_{xx} E)\rangle \,dt = & \int_0^T\langle \partial_x(F.\partial_x \delta E) -(\partial_x F).(\partial_x \delta E) \rangle \, dt\nonumber \\
= & \int_0^T\langle \partial_x(F.\partial_x \delta E)- \partial_x \big((\partial_x F).\delta E\big) 
+ \delta E .\partial_{xx} F\rangle \, dt\nonumber \\
= & \int_0^T\big[F.\partial_x \delta E  - (\partial_x F).\delta E \big]_{x=0}^{x=L}
+ \langle \delta E .\partial_{xx} F\rangle \, dt \nonumber \\
= & \int_0^T\langle \delta E .\partial_{xx} F\rangle \, dt.
\end{align}
The last step is achieved by applying the  boundary conditions 
\begin{align}
F(L,t)=&F(0,t) \\
\bar{N}(L,t)=&\bar{N}(0,t) \\
\tilde{N}^*(L,t)=&\tilde{N}^*(0,t).e^{+itSL} \\
\tilde{\Phi}^*(L,t)=&\tilde{\Phi}^*(0,t).e^{+itSL}. 
\end{align}
Similarly,
\begin{align}
\int_0^T\langle \tilde{\Phi}^*.\delta(\partial_{xx} \tilde{\phi})\rangle \,dt &= \int_0^T\langle \delta \tilde{\phi} .\partial_{xx} \tilde{\Phi}^*\rangle \, dt ,\\
\int_0^T\langle \tilde{N}^*.\delta(\partial_{xx} \tilde{n})\rangle \,dt &= \int_0^T\langle \delta \tilde{n} .\partial_{xx} \tilde{N}^*\rangle \, dt.
\end{align}
The terms that incorporate the background shear give us
\begin{align}
\int_0^T\langle \tilde{N}^*.\delta\big(i\tilde{n}(Sx+E)\big)\rangle\,dt 
&= \int_0^T\langle \delta \tilde{n}.\big(i\tilde{N}^*(Sx+E)\big)\rangle\,dt 
+\int_0^T\langle \delta E.\big(i\tilde{n}\tilde{N}^*\big)\rangle\,dt , \\
\int_0^T\langle \tilde{\Phi}^*.\delta\big(i\tilde{\phi}(Sx+E)\big)\rangle\,dt 
&= \int_0^T\langle \delta \tilde{\phi}.\big(i\tilde{\Phi}^*(Sx+E)\big)\rangle\,dt 
+\int_0^T\langle \delta E.\big(i\tilde{\phi}\tilde{\Phi}^*\big)\rangle\,dt.
\end{align}
Next,
\begin{align}
\int_0^T \langle \tilde{N}^*.\delta(i\tilde{\phi}\partial_x\bar{n})\rangle &=
\int_0^T \langle \delta\tilde{\phi}.(i\tilde{N}^*\partial_x\bar{n})\rangle \,dt
+\int_0^T \langle i\tilde{\phi}\tilde{N}^*\partial_x(\delta \bar{n})\rangle \,dt \nonumber \\
&= \int_0^T \langle \delta\tilde{\phi}.(i\tilde{N}^*\partial_x\bar{n})\rangle \,dt
+\int_0^T \langle \partial_x(i\tilde{\phi}\tilde{N}^* \delta \bar{n}) 
-  \delta \bar{n}\partial_x(i\tilde{\phi}\tilde{N}^*) \rangle \,dt \nonumber \\
&= \int_0^T \langle \delta\tilde{\phi}.(i\tilde{N}^*\partial_x\bar{n})\rangle \,dt
-\int_0^T \langle  \delta \bar{n}\partial_x(i\tilde{\phi}\tilde{N}^*) \rangle \,dt, 
\end{align}
having again applied boundary conditions. We now consider
\begin{align}
\int_0^T\langle \bar{N}. \delta\big(i\partial_x(\tilde{n}^*\tilde{\phi}-\tilde{n}\tilde{\phi}^*)\big)\rangle \,dt
=& \int_0^T\langle \partial_x\big( i\bar{N} \delta (\tilde{n}^*\tilde{\phi}-\tilde{n}\tilde{\phi}^*)\big)\rangle \,dt
-\int_0^T\langle i \delta (\tilde{n}^*\tilde{\phi}-\tilde{n}\tilde{\phi}^*)\partial_x\bar{N}\rangle \,dt \nonumber \\
=&-\int_0^T\langle \delta \tilde{n}^*.(i\tilde{\phi}\partial_x\bar{N})\rangle\,dt 
-\int_0^T\langle \delta \tilde{\phi}.(i\tilde{n}^*\partial_x\bar{N})\rangle\,dt \nonumber \\
&\qquad +\int_0^T\langle \delta \tilde{n}.(i\tilde{\phi}^*\partial_x\bar{N})\rangle\,dt 
+\int_0^T\langle \delta \tilde{\phi}^*.(i\tilde{n}\partial_x\bar{N})\rangle\,dt \nonumber \\
=& \int_0^T\langle \delta \tilde{n} .(2i\tilde{\phi}\partial_x\bar{N})\rangle \,dt 
- \int_0^T\langle \delta \tilde{\phi} .(2i\tilde{n}\partial_x\bar{N})\rangle \,dt.
\end{align}
The final term is
\begin{align}
\int_0^T\langle F. \delta\big(i\partial_x(\tilde{\phi}\partial_x\tilde{\phi}^*
-\tilde{\phi}^*\partial_x\tilde{\phi})\big)\rangle \,dt
=& \int_0^T\langle \partial_x\big(iF\delta(\tilde{\phi}\partial_x\tilde{\phi}^*
-\tilde{\phi}^*\partial_x\tilde{\phi})\big)\rangle \,dt
-\int_0^T\langle i\delta(\tilde{\phi}\partial_x\tilde{\phi}^*
-\tilde{\phi}^*\partial_x\tilde{\phi})\partial_x F \rangle \,dt \nonumber \\
=&\int_0^T \langle \delta\tilde{\phi}. (i\partial_x \tilde{\phi}^*\partial_xF)-
\delta\tilde{\phi}^*. (i\partial_x \tilde{\phi}\partial_xF)\rangle \,dt
+\int_0^T \langle i \tilde{\phi} \partial_x (\delta \tilde{\phi}^*)\partial_xF-
i \tilde{\phi}^* \partial_x( \delta \tilde{\phi})\partial_xF\rangle \,dt \nonumber \\
=&\int_0^T \langle \delta\tilde{\phi} .(i\partial_x \tilde{\phi}^*\partial_xF)-
\delta\tilde{\phi}^* .(i\partial_x \tilde{\phi}\partial_xF)\rangle \,dt
-\int_0^T \langle \delta\tilde{\phi}^*. \big(i\partial_x (\tilde{\phi}\partial_xF)\big)-
\delta\tilde{\phi} .\big(i\partial_x (\tilde{\phi}^*\partial_xF)\big)\rangle \,dt \nonumber \\
=&\int_0^T\langle \delta \tilde{\phi}.\big(2i\partial_x \tilde{\phi}^*\partial_xF)+
2i\partial_x(\tilde{\phi}^*\partial_xF)\big)\rangle\,dt.
\end{align}
All the remaining terms are trivial. Combining all of these together, and collecting terms yields
\begin{align}
\delta \mathscr{L}=&\langle \delta\bar{n}(x,T).\big(\bar{n}(x,T)-\bar{N}(x,T)\big)
+\delta E(x,T).\big(E(x,T)-F(x,T)\big) \nonumber \\
&\qquad \qquad \qquad +\delta\tilde{\phi}(x,T).\big(\tilde{\phi}(x,T)-\tilde{\Phi}(x,T)\big)
+\delta\tilde{n}(x,T).\big(\tilde{n}(x,T)-\tilde{N}(x,T)\big) \rangle \nonumber \\ 
&-\delta \lambda.\big(\langle \bar{n}(x,0)^2+E(x,0)^2+\tilde{\phi}(x,0)^2+\tilde{n}(x,0)^2 \rangle -A_0\big)\nonumber \\
&+\langle \delta\bar{n}(x,0).\big(\bar{N}(x,0)-\lambda \bar{n}(x,0)\big)
+\delta E(x,0).\big(F(x,0)-\lambda E(x,0)\big) \nonumber \\
&\qquad \qquad \qquad +\delta\tilde{\phi}(x,0).\big(\tilde{\Phi}(x,0)-\lambda \tilde{\phi}(x,0)\big)
+\delta\tilde{n}(x,0).\big(\tilde{N}(x,0)-\lambda\tilde{n}(x,0)\big) \rangle \nonumber \\ 
&-\int_0^T \langle \delta\bar{N}.\big(\partial_t\bar{n}-i\partial_x(\tilde{n}^*\tilde{\phi}-\tilde{n}\tilde{\phi}^*)\big) \rangle\,dt\nonumber  \\
&-\int_0^T \langle \delta F.\big(\partial_tE-i\partial_x(\tilde{\phi}\partial_x\tilde{\phi}^*-\tilde{\phi}^*\partial_x\tilde{\phi})-\partial_{xx}E\big) \rangle\,dt \nonumber \\
&-\int_0^T\langle \delta\tilde{\Phi}^*.\big(\partial_t\tilde{\phi}+i\tilde{\phi}(Sx+E) -i\tilde{n}-\partial_{xx}\tilde{\phi}\big) \rangle\,dt \nonumber \\
&-\int_0^T\langle \delta\tilde{N}^*.\big(\partial_t\tilde{n}+i\tilde{n}(Sx+E) -i\tilde{\phi}(\partial_x \bar{n} -1) -\partial_{xx}\tilde{n}\big) \rangle\,dt \nonumber \\
&-\int_0^T\langle\delta \bar{n}.\big(\partial_t\bar{N}-\frac{i}{2}\partial_x(\tilde{N}^*\tilde{\Phi}-\tilde{N}\tilde{\Phi}^*) \big) \rangle \,dt  \nonumber \\
&-\int_0^T\langle \delta E.\big(\partial_tF-\frac{i}{2}(\tilde{n}\tilde{N}^*-\tilde{n}^*\tilde{N})\big) \rangle \, \nonumber \\
&-\int_0^T\langle \delta \tilde{\phi}.\big(\partial_t\tilde{\Phi}+i\tilde{\Phi}(Sx+E) -i\tilde{N}(\partial_x \bar{n} -1) +2i\tilde{n}\partial_x\bar{N}
+2i\partial_x(\tilde\phi\partial_xF)+2i(\partial_x\tilde\phi)(\partial_xF)+\partial_{xx}\tilde{\Phi} \big) \rangle \,dt \nonumber \\
&-\int_0^T\langle \delta \tilde{n}.\big(\partial_t\tilde{N}+i\tilde{N}(Sx+E) -i\tilde{\Phi} -2i\tilde{\phi}\partial_x\bar{N}+\partial_{xx}\tilde{N} \big) \rangle \,dt.
\end{align}

This is only zero when all of the terms are individually zero. The first three sets of angled brackets give us the 
link between the direct and adjoint variables at time $T$, the requirement that the initial amplitude is fixed and 
gradient of the functional with respect to the initial condition, $\delta \mathscr{L}/\delta \mathbf{x}_0$. The first
four time integrals give us $\delta \mathscr{L}/\delta \bar{N}$,
$\delta \mathscr{L}/\delta F$, $\delta \mathscr{L}/\delta \tilde{\Phi}$ and $\delta \mathscr{L}/\delta \tilde{N}$ 
each of which is zero if the disturbance evolves according to the equations of motion for the system. The final 
four time integrals give us $\delta \mathscr{L}/\delta \bar{n}$, 
$\delta \mathscr{L}/\delta E$, $\delta \mathscr{L}/\delta \tilde{\phi}$ and $\delta \mathscr{L}/\delta \tilde{n}$. 
These are zero so long as the adjoint variables evolve according to the our four new equations of motion.

\end{document}